\documentclass[twocolumn,showpacs,preprintnumbers,amsmath,amssymb]{revtex4}
\usepackage{graphicx}    
\usepackage{dcolumn}    
\usepackage{bm}              
\usepackage{amsmath,amsfonts,amssymb}
\usepackage{graphicx,epic,eepic}
\usepackage{subfigure}
\usepackage{graphics}
\usepackage[dvips]{color}
\usepackage{epsfig}
\usepackage{units}      
\usepackage{mathrsfs}   
\newcommand{\reg}[1]{\hbox{\b{\tiny #1}}}
\newcommand{\Reg}[1]{\hbox{\b{#1}}}
\newcommand{\regreg}[3]{\hbox{\b{\tiny #1} {\tiny #2} \b{\tiny #3}}}
\newcommand{\mf}[1]{u^{\reg{#1}}_{k} }
\newcommand{\mfc}[1]{ {u^{\reg{#1}}_{k}}^{*} }
\newcommand{\mfE}[1]{v^{\reg{#1}}_{k} }
\newcommand{\mfEc}[1]{ {v^{\reg{#1}}_{k}}^{*} }
\newcommand{\mfqp}[1]{\tilde{u}^{\reg{#1}}_{k} }
\newcommand{\Mf}[1]{\vec{U}^{\reg{#1}}_{k}}
\newcommand{\MfE}[1]{\vec{V}^{\reg{#1}}_{k}}
\newcommand{\sound}[1]{c_{0}^{\reg{#1}} }
\newcommand{\vstate}[1]{\vert 0^{\reg{#1}} \rangle}
\newcommand{\vstatebra}[1]{\langle 0^{\reg{#1}} \vert}
\newcommand{\aop}[1]{{\hat{a}_{k}^{\reg{#1}}}}
\newcommand{\aopneg}[1]{{\hat{a}_{-k}^{\reg{#1}}}}
\newcommand{\aopdag}[1]{{\hat{a}_{k}^{\reg{#1}^{{\small  \dag}}}}}
\newcommand{\aopdagneg}[1]{{\hat{a}_{-k}^{\reg{#1}^{{\small  \dag}}}}}
\newcommand{\Ba}[2]{\alpha_{k}^{\reg{#1}\reg{#2}}}
\newcommand{\Bb}[2]{\beta_{k}^{\reg{#1}\reg{#2}}}
\newcommand{\Bac}[2]{{\alpha_{k}^{\reg{#1}\reg{#2}}}^{*}}
\newcommand{\Bbc}[2]{{\beta_{k}^{\reg{#1}\reg{#2}}}^{*}}
\def\tx{(t,\mathbf{x})}
\def\txp{(t,\mathbf{x}')}

\newcommand{\comBig}[2]{\Big[ #1,#2 \Big]}
\newcommand{\abs}[1]{\left\vert #1 \right\vert}
\newcommand{\op}[1]{\hat{#1}}

\newcommand{\refb}[1]{(\ref{#1})}
\def\ie{{\emph{i.e.}}}
\def\eg{{\emph{e.g.}}}
\begin{document}
\preprint{gr-qc/0703117}
\title{Trans-Planckian physics and signature change events in Bose gas hydrodynamics}
\author{Silke Weinfurtner}
\email{Silke.Weinfurtner@mcs.vuw.ac.nz}
\affiliation{School of Mathematics, Statistics, and Computer Science,\\
Victoria University of Wellington, PO Box 600, Wellington, New
Zealand}
\author{Angela White}
\email{Angela.White@anu.edu.au}
\affiliation{Centre for Gravitational Physics, Department of Physics,\\
The Australian National University, Canberra ACT 0200, Australia}
\author{Matt Visser}
\email{Matt.Visser@mcs.vuw.ac.nz}
\affiliation{School of Mathematics, Statistics, and Computer Science, \; \\
Victoria University of Wellington, PO Box 600, Wellington, New Zealand}
\date{27 February 2007; \LaTeX-ed \today}
\begin{abstract}
We present an example of emergent spacetime as the hydrodynamic
limit of a more fundamental microscopic theory. The low-energy,
long-wavelength limit in our model is dominated by collective
variables that generate an effective Lorentzian metric. This
system naturally exhibits a microscopic mechanism allowing us to
perform controlled signature change between Lorentzian and
Riemannian geometries. We calculate the number of particles
produced from a finite-duration Euclidean-signature event, where
we take the position that to a good approximation the dynamics is
dominated by the evolution of the linearized perturbations, as
suggested by Calzetta and Hu [Phys. Rev. A {\bf 68} (2003)
043625]. We adapt the ideas presented by Dray {\emph{et al.}}
[Gen. Rel. Grav. {\bf 23} (1991) 967], such that  the
field and its canonical momentum are continuous at the signature-change event.\\
We investigate the interplay between the underlying microscopic
structure and the emergent gravitational field, focussing on its
impact on particle production in the ultraviolet regime. In
general, this can be thought of as the combination of
trans-Planckian physics and signature-change physics. Further we
investigate the possibility of using the proposed signature change
event as an amplifier for analogue ``cosmological particle
production'' in condensed matter experiments.
\end{abstract}
\maketitle
%
%
\section{Introduction and Motivation \label{sec:Motivation}}
%
%
Emergent spacetimes \cite{Hu:2005wu, Barcelo:2005ln,
Matt.-Visser:2002ot} allow us to approach the subject of curved
spacetime quantum field theory (CST-QFT) through the back door.
This has been demonstrated in detail by using ultra-cold
(non-relativistic), highly dilute Bose gases
\cite{Garay:2000ez,Garay:2001in,Barcelo:2001gt}. Under appropriate
conditions the fundamental microscopic theory can be replaced by a
classical mean-field, the Bose--Einstein condensate (BEC)
\cite{Castin:2001aa}. Collective excitations,  both classical and
quantum,  experience an effective spacetime whose entries are
purely macroscopic mean-field variables. The kinematic equations
for linearized perturbations --- neglecting back-reaction and
finite temperature effects --- are equivalent to covariant
minimally coupled scalar fields, with a d'Alembertian
\begin{equation}
\Delta \ldots ={1\over\sqrt{-g}} \, \partial_{a}(\sqrt{-g} \, g^{ab} \, \partial_{b} \ldots)
\end{equation}
 defined by an effective metric $g_{ab}$ \cite{Garay:2000ez,Garay:2001in,Barcelo:2001gt}.\\

Quantum field theory in curved spacetime is a good approximation
for semiclassical gravity --- at the level where back-reactions of
the quantum fields on the gravitational field are negligible
\cite{Birrell:1984aa}. As a consequence quantum effects in curved
spacetimes (CSTs) --- \eg, Hawking radiation
\cite{Unruh:1981bi,Visser:1993tk,Visser:1998gn}  and cosmological
particle production  --- do not require emergent Einstein gravity
\emph{per se}; the existence of an emergent spacetime, an
effective gravitational field $g_{ab}$, is sufficient. Of course
both systems, semiclassical gravity and any analogue model, must
involve some dynamics --- and so they will eventually diverge from
each other. The only possible loophole would be if Einstein
gravity were itself to be the ``hydrodynamics'' of some more
fundamental theory (of microscopic objects, for example strings,
molecules, or atoms). For our purposes a perfect match is not
required, and we refer the interested reader to
\cite{Sakharov:1968dy, Jacobson:1995eg, Barcelo:2001lu,
Visser:2002hf, Volovik:2003jn} for further details.

It has been shown that the repulsive or attractive nature of
atomic interaction in a Bose gas is directly related to the
signature of the low-energy emergent metric: Lorentzian $(-,+++)$
for repulsive interactions, Euclidean $(+,+++)$ for attractive
interactions \cite{Barcelo:2001gt}. In 2001 a BEC experiment
\cite{Donley:2001aa,Roberts:2001aa} was carried out that can be
viewed as the first analogue model experiment. By tuning through a
Feshbach resonance \cite{Inouye:1998aa} the atomic interactions
were driven into a weakly attractive regime, and triggered a
controlled condensate collapse. Two years later, the theoretical
work of Calzetta and Hu  \cite{Calzetta1:2003xb,Calzetta:2005yk},
connected the the so-called Bose-nova phenomenon with the
amplification, mixing of positive and negative modes, and
squeezing of vacuum fluctuations due to a signature change event.
Perhaps surprisingly, the calculations carried out in
\cite{Calzetta1:2003xb,Calzetta:2005yk} did not include background
condensate dynamics and yet their theoretical predictions reflect
the experimental data relatively well. For short time-intervals of
attractive atom-atom interaction (\ie, a brief excursion into
Euclidean signature), the Bose-nova event is dominated by the
evolution of the quantum perturbations, and to a good
approximation independent of the background condensate dynamics.

Specifically, given that we do not as yet have any precise
detailed  model for emergent Einstein gravity, as opposed to
emergent curved spacetime, it seems necessary to focus on quantum
effects that are merely of kinematic rather than dynamic nature.
However, there are ways to study the influence of possible quantum
gravity candidates with CST-QFT. This branch of physics is called
quantum gravity phenomenology (QGP) \cite{Mattingly:2005aa}.
Emergent spacetimes can be used to analyze some portions of QGP,
where Lorentz invariance violations (LIV) are present at
ultraviolet scales. The LIV scale is supposed to be connected with
the Planck length, where new physics is expected. This idea is
naturally implemented in any emergent spacetime model. For
example, the BEC-based analogue models only recover Lorentz
invariance (LI) for the low-energy, long-wavelength phonon modes.
For trans-phononic modes microscopic corrections (\eg, quantum
pressure effects) have to be absorbed into the macroscopic
picture. The borderline between the two modes can be viewed as the
analogue LIV scale. Given that trans-phononic modes start to see
first signs of the fundamental microscopic theory, it makes sense
to speak of LIV at the analogue Planck scale. The key reason why
it is interesting to study modifications in the dispersion
relation is that many different effective field theories (EFTs)
already predict deviations at the kinematical level. A detailed
treatment of the analogue trans-Planckian model can be found in
\cite{Visser:2001ix,Liberati:2006sj,Liberati:2006kw,Weinfurtner:2006iv,Weinfurtner:2006nl}.

In ongoing work  \cite{Jain:2006ki} a classical phase-space method has
been used to numerically simulate cosmological particle production in BECs.
There it is shown that for a consistent treatment microscopic corrections
play an important role in the emergent spacetime picture. This leads to
emergent ``rainbow metrics'', with a Planck-suppressed momentum dependence
for the modes, and consequently leads to a modification in the quasi-particle
spectrum.

Our central idea in the current article is to merge all of the
above, to address the trans-Planckian problem for a signature
change event in a Bose gas. We are particularly interested in the
ultraviolet physics of the phonon modes and hence have chosen a
specific BEC set-up where the external trapping potential does not
interfere with the dynamics (\eg, hard-walled box). Step-by-step,
in section  \refb{sec:BEC.CST-QFT} we show how spacetime emerges
from a Bose gas; calculate the quasi-particle production from
sudden sound speed variations in section
\refb{sec:Sudden.Lorentzian.Variations}; extend this calculation
to sudden variations for finite regions with different signature
(Lorentzian $\leftrightarrow$ Euclidean) in section
\refb{sec:Sudden.Euclidean.Variations}; introduce ultraviolet
physics and re-calculate the quasi-particle production in section
\refb{sec:UV-corrections}; suggest in section \refb{sec:amplifier} to employing a finite Euclidean region as a particle amplifier for cosmological particle production in a BEC; and last but certainly not least, we address the theoretical and experimental impact of our results in section \refb{sec:EndHappy}.
%
\section{Emergent geometry from a Bose gas \label{sec:BEC.CST-QFT}}
%
The intent in this section is to give readers unfamiliar with the topic some
understanding of analogue spacetimes. Following the example of
\cite{Barcelo:2005ln} we introduce an ultra-cold (\ie, non-relativistic),
highly dilute and weakly interacting gas of Bosons, using the formalism of
canonical quantization, and derive the equation of motion for small quantum
fluctuations around some classical background; better known as a BEC, see
section \refb{subsec:Bose-gas}. The Bose--Einstein condensate is a state of
matter where the Bosons macroscopically occupy the lowest quantum state. In
section \refb{subsec:A-spacetime} we focus on the hydrodynamic case and recover
a covariant minimally coupled massless scalar field for small quantum
fluctuations in the BEC.
Subsequently, in section \refb{sec:UV-corrections}, we will revisit this
derivation and include trans-phononic modes (\eg, ultraviolet physics) into the
emergent spacetime picture.
%
\subsection{Ultra-cold, weakly interacting Bose gas\label{subsec:Bose-gas}}
Suppose we have a system of $N$ Bosons. In quantum field theory the field variables are quantum operators that act on quantum states (Hilbert space of states; \eg, Fock space).
Field operators either create, $\op{\psi}^{\dag}\tx$, or destroy, $\op{\psi}\tx$, an individual Boson at a particular point in space and time, and satisfy the commutators:
\begin{eqnarray}
\comBig{\op{\psi}\tx }{\op{\psi}\txp}=\comBig{\op{\psi}^{\dag}\tx}{\op{\psi}^{\dag}\txp}=0\,;
\label{Eq:com1}\\
\comBig{\op{\psi}\tx}{\op{\psi}^{\dag}\txp}=\delta(\mathbf{x}-\mathbf{x}')\,.
\label{Eq:com2}
\end{eqnarray}
For a gas of trapped, ultra-cold, highly dilute and weakly interacting Bosons the Hamiltonian is given by
\begin{eqnarray} \label{Eq:Hamiltonian}
\op{H}= \int{ d\mathrm{x} \left(
- \op{\psi}^{\dag} \frac{\hbar^{2}}{2m}\nabla^{2} \op{\psi}
+ \op{\psi}^{\dag} V_{\mathrm{ext}} \op{\psi}
+ \frac{U}{2} \, \op{\psi}^{\dag}\op{\psi}^{\dag}\op{\psi}\op{\psi} \right) } \, .
\end{eqnarray}
This is a sum of  the kinetic energy of the Boson field, and the
two potential energy contributions; the external trap
$V_{\mathrm{ext}}$, and the particle interactions. The extreme
dilution of the gas (\eg, $10^{13}-10^{15}\,
\mathrm{atoms}/\mathrm{cm}^{3}$) suppresses more-than-two-particle
interactions, and in the weakly interacting regime the actual
inter-atom potential has been approximated by a pseudo-contact
potential,
\begin{equation} \label{Eq:U}
U = \frac{4 \pi \hbar^{2} a}{m}\,.
\end{equation}
Here $m$ is the single-Boson mass, and $a$ the $s$-wave scattering length. The sign of the scattering length determines the qualitative behavior of the interactions,
\begin{equation} \label{Eq:nature_of_a}
\begin{array}{rl}
a > 0 & \quad\mathrm{repulsive}\, ; \\
a < 0 & \quad \mathrm{attractive}\, .
\end{array}
\end{equation}
Negative and positive values of $a$ are experimentally accessible by tuning external magnetic fields, that interact with the inter-atomic potential; this process is called Fesh\-bach resonance \cite{Inouye:1998aa}. We would like to emphasize the importance of Eq.~\refb{Eq:nature_of_a} for the remaining sections.

We now have all the necessary information about our system, encoded in Eqs.~\refb{Eq:com1}--\refb{Eq:Hamiltonian},  to calculate its dynamics. We use the Heisenberg equation of motion to get the time-evolution for the field operator;
\begin{equation} \label{Heisenberg}
i\hbar \frac{\partial \op{\psi}}{\partial t} = \left[ -\frac{\hbar^{2}}{2m} \nabla^{2} + V_{\mathrm{ext}} + U\, \op{\psi}^{\dag}\op{\psi}  \right] \op{\psi} \,.
\end{equation}
To apply this discussion to the emergent spacetime programme, we use the macroscopic occupation of the lowest quantum state below a critical temperature (\eg, for alkali gases, below $10^{-5}$ K). If the cooling process prohibits the gas to solidify, a new state of matter will occur, the Bose--Einstein condensate. The condensate is a complex-valued macroscopic mean-field
\begin{equation} \label{Eq:psi}
\langle \op{\psi}\tx \rangle =
\psi\tx = \sqrt{n_{0}\tx} \exp\left(i\phi_{0}\tx\right)\,,
\end{equation}
where the individual microscopic particles give way to collective variables. We will show that the condensate density, $n_{0}$, and phase, $\phi_{0}$, define the analogue spacetime for small quantum fluctuations in the BEC. The essential step in deriving the kinematics for the perturbation is to separate the perturbation  from the condensate:
\begin{eqnarray}
\label{Eq:lin1}
\op{\psi} &\simeq& \psi\tx + \delta \op{\psi}\tx\,; \\
\label{Eq:lin2}
\op{\psi}^{\dag} &\simeq& \psi^{\dag}\tx + \delta \op{\psi}^{\dag}\tx\,.
\end{eqnarray}
This transformation is canonical if the creation $\delta\op{\psi}^{\dag}$ and destruction $\delta\op{\psi}$ of perturbations is consistent with the commutator Eqs.~\refb{Eq:com1}-\refb{Eq:com2}. Therefore, the commutators for the linearized quantum fluctuations are
\begin{eqnarray}
\comBig{\delta\op{\psi}\tx }{\delta\op{\psi}\txp}=\comBig{\delta\op{\psi}^{\dag}\tx}{\delta\op{\psi}^{\dag}\txp}=0\,;
\label{Eq:delta_com1}\\
\comBig{\delta\op{\psi}\tx}{\delta\op{\psi}^{\dag}\txp}=\delta(\mathbf{x}-\mathbf{x}')\,.
\label{Eq:delta_com2}
\end{eqnarray}

The quantum perturbations are small perturbations which, as per Eq.~\refb{Eq:psi}, are represented by two collective parameters; the density $n$, and the phase $\phi$. Clearly, any quantum perturbation should be related to variations in these two parameters; $n \simeq n_{0} + \op{n}$ and $\phi \simeq \phi_{0} + \op{\phi}$. A straightforward expansion of $\psi$ and $\psi^{\dag}$ around $n_{0}$ and $\phi_{0}$ leads to
\begin{eqnarray}
\label{Eq:exp.psi_1}
\delta\op{\psi} &\simeq& \psi \left( \frac{1}{2}\frac{\op{n}}{n_{0}} + i \op{\phi}  \right)\,, \\
\label{Eq:exp.psi_2}
\delta\op{\psi}^{\dag} &\simeq& \psi^{*} \left( \frac{1}{2}\frac{\op{n}}{n_{0}} - i \op{\phi} \right)\,.
\end{eqnarray}
In this way Eqs.~\refb{Eq:lin1} and \refb{Eq:lin2} are
compatible with Eq.~\refb{Eq:psi}. Thus the density,
$\op{n}$, and phase, $\op{\phi}$, fluctuations operators are
Hermitian operators:
\begin{eqnarray}
\op{n} \simeq n_{0} \left[ \frac{\delta\op{\psi}}{\psi} + \left(\frac{\delta\op{\psi}}{\psi}\right)^{\dag}  \right]\,;\\
\op{\phi} \simeq - \frac{i}{2} \left[ \frac{\delta\op{\psi}}{\psi} - \left(\frac{\delta\op{\psi}}{\psi}\right)^{\dag}\right]\,.
\end{eqnarray}
It is easy to see that the new operators are a set of canonical variables:
\begin{eqnarray}
\comBig{\op{n}\tx }{\op{n}\txp}=\comBig{\op{\phi}\tx}{\op{\phi}\txp}=0\,;
\label{Eq:nn_psipsi_com1}\\
\comBig{\op{n}\tx}{\op{\phi}\txp}=i\delta(\mathbf{x}-\mathbf{x}')\,.
\label{Eq:n_psi_com2}
\end{eqnarray}
The latter can be further modified, and we will subsequently revisit Eq.~\refb{Eq:n_psi_com2}.
While the split into background plus perturbation is up to this
point exact, we now linearize by assuming the perturbation to be
small, allowing us to neglect quadratic and higher-order products
of the perturbation field $\delta \psi$. (There are also more
sophisticated calculational techniques available based on the
Hartree--Fock--Bogoliubov--Popov approximation, but they are an
unnecessary complication in the present situation.)

Merging equations (\ref{Eq:lin1})--(\ref{Eq:lin2}) with equations (\ref{Eq:exp.psi_1})--(\ref{Eq:exp.psi_2}), applying them to Eq.~\refb{Heisenberg}, we obtain two equations,
\begin{eqnarray}
\label{hydro.n.1}
\frac{\partial \hat n}{\partial t} + \nabla \left[ \frac{n_{0} \hbar}{m} \nabla \hat \phi + \hat n \cdot \mathbf{v} \right] = 0\,; \\
\label{hydro.n.2}
\frac{\partial \hat \phi}{\partial t} + \mathbf{v} \cdot \nabla \hat \phi + \frac{\mathcal{U}}{\hbar} \, \hat n = 0\,.
\end{eqnarray}
Here we introduce the background velocity of the condensate,
\begin{equation}
\mathbf{v} = \frac{\hbar}{m} \nabla \phi_{0}.
\end{equation}
The quantity $\mathcal{U}$ can be thought of as an effective atomic interaction, as seen by the collective excitations. For long-wavelength, low-energetic modes this simplifies to Eq.~\refb{Eq:U}, the usual pseudo-contact potential;
\begin{equation} \label{Eq.hydro.approx}
\mathcal{U} \rightarrow U \quad : \quad \mathrm{phononic}\; \mathrm{modes}\,.
\end{equation}
We will come back to this point in section \refb{sec:UV-corrections} where we shall write down the expression for $\mathcal{U}$ in general, and in the eikonal limit, in a manner appropriate to describe trans-phononic (or, ultraviolet) modes. In the language of condensed matter physics, we include ``quantum pressure'' effects. Without quantum pressure we restrict our analysis to the phononic regime, and assume that all collective excitations propagate with the same speed, the sound speed $c_{0}$;
\begin{equation} \label{Eq:sound.speed.hydro}
c_{0}^{2} = \frac{n_{0}U}{m}\,.
\end{equation}

Before we continue with our program, we would like to revisit the
commutator derived in Eq.~\refb{Eq:n_psi_com2}. In the
phononic regime, we are able to write Eq.~\refb{hydro.n.2} as
follows,
\begin{equation} \label{Eq:n.explicit}
\hat n = - \frac{\hbar}{U} \left( \frac{\partial}{\partial_{t}} + \mathbf{v} \cdot \nabla \right) \hat \phi
= -\frac{\hbar}{U} \frac{D\hat \phi}{Dt} \, .
\end{equation}
Thus the commutator \refb{Eq:n_psi_com2} can be written in terms of $\hat \phi$ and its fluid-following derivative (or material derivative) $D\hat\phi / Dt$:
\begin{equation} \label{Eq:Pi_psi_com2}
\comBig{\hat\phi\tx}{\hat \Pi_{\hat \phi}\txp}=i \delta(\mathbf{x}-\mathbf{x}')\,.
\end{equation}
Here we have defined,
\begin{equation} \label{Eq:conj.monmentum.1}
\hat \Pi_{\hat \phi} = \frac{\hbar}{U}  \frac{D\hat\phi}{Dt}\,,
\end{equation}
which can be viewed as the conjugate momentum of $\hat \phi$; see Eq.~\refb{Eq:emergent.Lagrange.density} below.

The relation \refb{Eq:n.explicit} enables us to read density
perturbations as fluid-following derivatives of the phase
perturbations, and hence supplies us with the necessary tool to
eliminate all occurrences of $\hat n$ from equation
\refb{hydro.n.1}:
\begin{eqnarray}
\label{Eq:phase.evolution}
&&-\partial_{t} \left[ \frac{\hbar}{U} \, \partial_{t} \hat \phi \right]
-\partial_{t} \left[ \frac{\hbar}{U} \, \mathbf{v} \, \nabla \hat \phi \right]
-\nabla \left[ \frac{\hbar}{U} \, \mathbf{v} \, \partial_{t} \hat \phi  \right]
\nonumber\\
&& \qquad
+\nabla \left[ \frac{n_{0} \hbar}{m} \, \nabla \hat \phi -\frac{\hbar}{m} \, \mathbf{v} \left(\nabla\hat \phi\right) \mathbf{v} \right] =0 \,.
\end{eqnarray}
This equation, governing the kinematics for the phase perturbations $\hat \phi$, is the connection between condensed matter physics and emergent quantum field theory in curved spacetimes.

\subsection{Analogue spacetime \label{subsec:A-spacetime}}
A  compact and insightful way to express the evolution of phase
perturbations \refb{Eq:phase.evolution} is
\begin{equation} \label{Eq:KLG.1}
\frac{1}{\sqrt{\left\vert \det(g_{ab}) \right\vert}} \; \partial_{a} \left( \sqrt{ \left\vert \det(g_{ab}) \right\vert }\, g^{ab} \, \partial_{b}  \right) \hat \phi =0\;, \\
\end{equation}
where we introduce
\begin{equation} \label{Eq:g.1}
g_{ab} = \left( \frac{c_{0}}{U/\hbar} \right)^{\frac{2}{d-1}}
\left[
\begin{array}{cccc}
-\left(c_{0}^{2}-\mathbf{v}^{2}\right) & -v_{x} & -v_{y} & -v_{z} \\
-v_{x} & 1 & 0 & 0 \\
-v_{y} & 0 & 1 & 0 \\
-v_{z} & 0 & 0 & 1
\end{array}
\right]\, ;
\end{equation}
a covariant metric rank two tensor, whose entries are purely collective variables. The conformal factor depends on the spatial dimensionality, $d$, of the condensate cloud.
To derive Eq.~\refb{Eq:KLG.1}, we first write Eq.~\refb{Eq:phase.evolution} as,
\begin{equation} \label{Eq:KLG.f}
\partial_{a} \left( f^{ab} \; \partial_{b} \hat \phi \right)=0\,,
\end{equation}
where $f^{ab}$ is easily found to be,
\begin{equation} \label{Eq:f.1}
f^{ab} :=
\left[
\begin{array}{cccc}
-\frac{\hbar}{U} & -\frac{\hbar}{U} v_{x} & -\frac{\hbar}{U} v_{y} & -\frac{\hbar}{U} v_{z} \\
-\frac{\hbar}{U} v_{x} &  \frac{n_{0}\hbar}{m}-\frac{\hbar}{U}v_{x}^{2} &  -\frac{\hbar}{U}v_{x}v_{y} & -\frac{\hbar}{U}v_{x}v_{z} \\
-\frac{\hbar}{U} v_{y} & -\frac{\hbar}{U}v_{x}v_{y} & \frac{n_{0}\hbar}{m}-\frac{\hbar}{U}v_{y}^{2} & -\frac{\hbar}{U}v_{y}v_{z} \\
-\frac{\hbar}{U} v_{z} & -\frac{\hbar}{U}v_{x}v_{z} & -\frac{\hbar}{U}v_{y}v_{z} & \frac{n_{0}\hbar}{m}-\frac{\hbar}{U}v_{z}^{2} \\
\end{array}
\right]\,.
\end{equation}
The two equations, \refb{Eq:KLG.f}, and \refb{Eq:KLG.1}, are equivalent if
\begin{equation}
f^{ab} =\sqrt{\left\vert \det(g_{ab}) \right\vert } \; g^{ab}.
\end{equation}
Here $g^{ab}$ is a contravariant tensor, and since $g^{ab}g_{cb}=\delta^{a}{}_{c}$, it is only a question of matrix inversion to find its covariant equivalent, \refb{Eq:g.1}. For considerably more details and a thorough derivation we suggest the following literature: \cite{Barcelo:2005ln,Visser:2005hs,Barcelo:2001gt,Garay:2001in}.

At this stage we would like to comment on the physical
implications of the results presented so far. The motivation to
write the differential Eq.~\refb{Eq:phase.evolution},
governing the  excitation spectrum in the form presented in
\refb{Eq:KLG.1}, is to find an analogy for the curved spacetime
Klein--Gordon equation which describes minimally coupled spin-zero
Bosons in curved spacetime.
It is appropriate to define an emergent Lagrange density,
\begin{equation} \label{Eq:emergent.Lagrange.density}
\mathcal{L}=-\frac{1}{2} f^{ab} \; \partial_{a}\hat\phi \; \partial_{b}\hat\phi\, ,
\end{equation}
such that Eq.~\refb{Eq:KLG.1} can be obtained as the
Euler-Lagrange equations justified by the principle of least
action. The momentum conjugate to $\hat \phi$ is specified by
\begin{equation}
\Pi_{\hat\phi} := {\partial \mathcal{L}\over\partial (\partial_{t} \hat\phi)}
=-f^{tb} \; \partial_{b}\hat\phi,
\end{equation}
and hence is in agreement with equation
\refb{Eq:conj.monmentum.1}. We see that quantum phase and density
perturbations in a Bose--Einstein condensate are a canonical set
of field and conjugate field operators on the emergent spacetime.

We now temporarily set aside this analogy and apply our model to particle production in non-smooth emergent geometries.
%
\section{Sudden changes in spacetime geometry \label{sec:Sudden.Lorentzian.Variations}}
%
In this section we calculate the mixing of positive and negative frequencies due to ``sudden'' step-wise variations in the sound speed, Eq.~\refb{Eq:sound.speed.hydro}. The initial and final emergent geometries are now flat Minkowski spacetimes, which are discontinuously connected at the step. Physically the step is generated by a very rapid change in the magnetic field, which very rapidly drives one through a Feshbach resonance, which in turn very rapidly changes the scattering length $a$, and so finally induces a  rapid change in the speed of sound.

\subsection{Quantum fields as harmonic oscillators \label{subsec:Label}}
An idealized Bose--Einstein condensate trapped in a finite quantization box of volume $L^{d}$
is comparable to flat Minkowski spacetime. Here the macroscopic parameters are zero background velocity, $\mathbf{v}=\mathbf{0}$, and constant sound speed \refb{Eq:sound.speed.hydro}. That is, one considers a uniform number density, $n_{0}\tx = \mathrm{const}$, and a fixed scattering length, $a(t)=\mathrm{const}$. Therefore the emergent metric given in \refb{Eq:g.1} is a diagonal tensor whose entries are time- and space independent;
\begin{equation} \label{Eq:eta.1}
f^{ab} =
\left[
\begin{array}{cccc}
-\frac{\hbar}{U} & 0 & 0 & 0 \\
0 & \frac{n_{0} \hbar}{m} & 0 & 0 \\
0 & 0 & \frac{n_{0} \hbar}{m} & 0 \\
0 & 0 & 0 & \frac{n_{0} \hbar}{m}
\end{array}
\right]\, .
\end{equation}

We employ the canonical variables on our effective relativistic spacetime to write the Klein--Gordon Eq.~\refb{Eq:KLG.1} as
\begin{equation} \label{Eq:KLG.2}
\partial_{t} \left( \frac{\hbar}{U} \; \partial_{t} \hat \phi   \right) - \frac{n_{0} \hbar}{m} \nabla^{2} \hat\phi = 0\,.
\end{equation}
It is possible to decouple the Klein--Gordon field \label{Eq:KLG.2} into independent Harmonic oscillators. To show this, we make use of
\begin{equation}
\hat \phi\tx = \frac{1}{L^{d/2}}\sum_{k}{ \frac{1}{\sqrt{2}}
\left[ \hat a_{k} v_{k}^{*}(t) e^{i\mathbf{kx}}+\hat a_{k}^{\dag} v_{k}(t) e^{-i\mathbf{kx}} \right]}\, .
\end{equation}
Note that for a hard-walled box the modes fulfill non-periodic boundary conditions.
In Minkowski spacetime there exists a natural set of mode functions,
\begin{equation} \label{Eqn:Mode.Functions.Lorentzian}
v_{k}(t) =\sqrt{\frac{U}{\hbar}} \, \frac{1}{\sqrt{2\omega_{k}}}
\, e^{i\omega_{k} t}\, ,
\end{equation}
associated with the Poincar\'e group, a symmetry group of the Minkowski line-element. Here $\partial_{t}$ is a time-translation Killing vector which can also be thought of as a differential operator with eigenvalues $(-i\omega_{k})$, where $\omega_{k}>0$ are said to be positive frequency modes. Hence the vacuum is invariant under the action of the Poincar\'e group and all observers agree on existence, or non-existence, of particles in flat spacetime. The physics is observer-independent, as expected. For a more detailed treatment see reference \cite{Birrell:1984aa}, and the appendix below.

The decoupled equations for the mode-operators are
\begin{equation} \label{Eq:KLG.3}
\dot{\hat a}_{k}(t) = - i \omega_{k} \, \hat a_{k}(t)\, ; \quad  \mathrm{and} \quad \dot{\hat a}_{k}^{\dag}(t) = + i \omega_{k} \, \hat a_{k}^{\dag}(t)\, ,
\end{equation}
where $\dot{\hat a}_{k}(t) = \hat a_{k} v_{k}^{*}(t)$ and
$\dot{\hat a}_{k}^{\dag}(t) = \hat a_{k}^{\dag} v_{k}(t)$. For
now, we are working in the low-energy, long-wavelength regime and
obtain a ``relativistic'' dispersion relation for the modes:
\begin{equation} \label{Eq.disp.rel.hydro}
\omega_{k} = c \, k\,,
\end{equation}
Later on, in section \refb{sec:UV-corrections}, we will include trans-phononic modes into our picture, and see how the microscopic structure induces LIV breaking terms for high-energy, short-wavelength perturbations.

The canonical creation and destruction operators obey the usual commutator
\begin{equation} \label{Eq:Pi_psi_com2.Minkowski.new}
\comBig{\hat a_{k}(t)}{\hat a_{-k'}^{\dag}(t)}= \delta_{kk'}\,,
\end{equation}
and acting on the particle basis for the Hilbert space of states, the Fock states $\vert \;\rangle$, they are a powerful tool to calculate the number of particles $n_{k}$ in the mode labeled by $k$.
The normalized basis vectors can be obtained from the vacuum, or zero-particle state, $\vert 0 \rangle$. Thus is the state that is destroyed by operators $\hat a_{k} \vert 0 \rangle = 0$, for all modes labeled by $k$.
In general
\begin{equation}
\hat a_{k}^{\dag} \vert n_{k} \rangle = (n+1)^{1/2} \vert (n+1)_{k}\rangle \, ,
\end{equation}
 and
\begin{equation}
\hat a_{k} \vert n_{k} \rangle = n^{1/2} \vert (n-1)_{k} \rangle \, ,
\end{equation}
so that we can define the number operator $\hat N_{k}$,
\begin{equation}
\hat N_{k} \vert n_{k} \rangle = \hat a_{k}^{\dag} a_{k} \vert n_{k} \rangle = n_{k} \vert n_{k} \rangle \, .
\end{equation}

We now use the tools we have presented above to calculate the
quasi-particle production in a BEC-based emergent spacetime due to
sudden changes in the sound speed; such that we patch two flat
spacetimes in a step-wise fashion, by suitably changing the
microscopic parameters.

\subsection{Particle production \label{subsec:Label}}
From the emergent spacetime point of view step-wise changes in the microscopic parameters induce sudden variations in the collective variables, and enforce discontinuous interchanges between different Minkowski spacetimes. In terms of our emergent spacetime from a Bose gas, this can be achieved through an external magnetic field, that adjusts the atomic interactions, and consequently the scattering length $a$ and the sound speeds $\sound{i}$.

Following the ideas of Dray \emph{et al.} \cite{Dray:2004aa,Dray:1991zz}, we assume that the fields are continuous,
\begin{equation} \label{Eq.C1}
\left. \hat\phi^{\reg{i}} - \hat\phi^{\reg{j}}  \right\vert_{_{\sum}}= 0 \, ,
\end{equation}
on the space like hypersurface {\small $\sum$} at  fixed time
$t=t^{\reg{i}\reg{j}}$. Here $t=t^{\reg{i}\reg{j}}$ is the time at
which a transition occurs from $c^{\reg{i}}$ to $c^{\reg{j}}$. The
Klein--Gordon equation in each region can be written in terms of
exterior derivatives,
\begin{equation}
(-1)^{n}\partial_{a} \left( f^{ab}\,\partial_{b} \hat\phi^{\reg{i}} \right)
= *d(*d\hat\phi^{\reg{i}})
=: *dF^{\reg{i}}
=0\, ;
\end{equation}
hence in each region $dF^{\reg{i}}=0$, where $F^{\reg{i}}$ is an exact $(n-1)$-form. These forms are connected discontinuously;
\begin{equation}
F = \Theta \, F^{\reg{i}} + (1-\Theta) F^{\reg{j}}\, ,
\end{equation}
where $\Theta$ is the usual Heavyside function. Thus, we get
\begin{eqnarray}
dF &=& \Theta dF^{\reg{i}} + (1-\Theta) dF^{\reg{j}} + \delta(t)\, dt\wedge [F]  \\
&=&  \delta(t)\, dt\wedge [F] \, ,
\end{eqnarray}
where $[F]=F{^{\reg{i}}}-F{^{\reg{j}}}$. This supplies us with a connection condition for the canonical momentum,
\begin{equation}
\delta(t)\; dt\wedge [F] =0\, ;
\end{equation}
\ie,
\begin{equation} \label{Eq.C2}
\left. \hat\Pi_{\hat \phi}{}^{\reg{i}} - \hat\Pi_{\hat \phi}{}^{\reg{j}}  \right\vert_{_{\sum}}= 0 \, .
\end{equation}
Thus the field and its canonical momentum must be continuously connected across the spatial hypersurface of sudden variation. In the following we apply the connection conditions we have just found, and calculate the number of quasi-particles produced by sudden sound speed variations in our BEC.

In each region $\Reg{i}$ the mode expansion,
\begin{equation}
\hat \phi^{\reg{i}} \tx = \frac{1}{\sqrt{2 L^{d}}} \sum_{\mathbf{k}} e^{i\mathbf{kx}} \left[ \mfc{i}(t) \aop{i} + \mf{i}(t) \aopdagneg{i} \right]
\end{equation}
involves a distinct set of creation $\aopdag{i}$ and destruction $\aop{i}$ operators. The mode functions $\mf{i}$ and $\mfc{i}$ obey the curved-spacetime Klein--Gordon equation,
\begin{equation} \label{Eq:KLG.4}
\partial_{t} \left[ \frac{\hbar}{U^{\reg{i}}} \; \partial_{t}  \mf{i}(t)   \right] - \frac{n_{0} \hbar}{m}\, k^{2} \; \mf{i}(t) = 0\,;
\end{equation}
they form a complete basis for the two dimensional solution space. At the transition time $t^{\reg{i}\reg{j}}$ we connect the mode functions and their first derivatives,
\begin{eqnarray}
\label{Eq.C1}
\Mf{i}(t^{\reg{i}\reg{j}}) &=& M^{\reg{i}\reg{j}} \; \Mf{j}(t^{\reg{i}\reg{j}})\, , \\
\label{Eq.C2}
\frac{\hbar}{U^{\reg{i}}}\; \partial_{t}\Mf{i}(t^{\reg{i}\reg{j}}) &=&
M^{\reg{i}\reg{j}} \; \frac{\hbar}{U^{\reg{j}}} \; \partial_{t}\Mf{j}(t^{\reg{i}\reg{j}}) \,.
\end{eqnarray}
Here we have combined both mode functions into one mode vector $\Mf{i}$,
\begin{equation}
\Mf{i}(t) =\left( \begin{array}{l} \mf{i}(t) \\ \mfc{i}(t) \end{array} \right)\, .
\end{equation}
In the cases we are interested in the transition matrix $M^{\reg{i}\reg{j}} = M(t^{\reg{i}\reg{j}})$ is time-independent, and for a pair of complex conjugate mode functions in each region $\Reg{i}$ and $\Reg{j}$, the matrix is of the form
\begin{equation} \label{Eq.M}
M^{\reg{i}\reg{j}} = \left( \begin{array}{cc}
\Ba{i}{j} & \Bb{i}{j} \\ \Bbc{i}{j} & \Bac{i}{j}
\end{array} \right)\, .
\end{equation}
The Wronskian $W[\mf{i},\mfc{i}]=(\partial_{t}\mf{i})\mfc{i}-\mfc{i}(\partial_{t}\mfc{i})$ of the mode functions is time-independent, see Eq.~\refb{Eq:KLG.4}, which implies a constraint on the Bogliubov coefficients $\Ba{i}{j}$ and $\Bb{i}{j}$;
\begin{equation} \label{Eq.KG.condition}
\vert \Ba{i}{j} \vert^{2} - \vert \Bb{i}{j} \vert^{2} = 1\, .
\end{equation}
To start with, consider a no-particle state in the initial Minkowski spacetime (region $\Reg{i}$), such that $\aop{i}\vstate{i}=0$. The correlation between the mode operators can easily be constructed from $\hat \phi^{\reg{i}}(t^{\reg{i}\reg{j}})=\hat \phi^{\reg{j}}(t^{\reg{i}\reg{j}})$, considering Eq.~\refb{Eq.C1} and Eq.~\refb{Eq.KG.condition} we get
\begin{eqnarray}
\aop{i} &=& \Ba{i}{j} \; \aop{j} - \Bb{i}{j} \; \aopdagneg{j}\, , \\
\aopdag{i} &=& \Bac{i}{j} \; \aopdag{j} - \Bbc{i}{j} \; \aopneg{j} \, .
\end{eqnarray}
Consequently, the mean number of $\Reg{i}$-particles in the $\Reg{j}$-vacuum,
\begin{equation}
\vstatebra{j} \hat N_{k}^{i} \vstate{j} = \vert \Bb{i}{j} \vert^{2} \; \delta^{d}(0)\, ,
\end{equation}
depends only on $|\Bb{i}{j}|$.

Explicitly, the elements for the transition matrix in a sudden change from region $\Reg{i}$ to region $\Reg{j}$ at a time $t^{\reg{i}\reg{j}}$ are given by:
\begin{eqnarray}
\label{Eq.alpha.one.step}
\Ba{i}{j} = \frac{1}{2} \left( \sqrt{X^{\reg{i}\reg{j}}} + \frac{1}{\sqrt{X^{\reg{i}\reg{j}}}} \right) \, e^{i (+\omega_{k}^{\reg{i}} - \omega_{k}^{\reg{j}})t^{\reg{i}\reg{j}}}\, , \\
\label{Eq.beta.one.step} \Bb{i}{j} = -\frac{1}{2} \left(
\sqrt{X^{\reg{i}\reg{j}}} - \frac{1}{\sqrt{X^{\reg{i}\reg{j}}}}
\right) \, e^{i (+\omega_{k}^{\reg{i}} +
\omega_{k}^{\reg{j}})t^{\reg{i}\reg{j}}} \, ,
\end{eqnarray}
where to simplify formulae it is convenient to introduce the ratio of change in the dispersion relations $X^{\reg{i}\reg{j}}$,
\begin{equation} \label{Eq.X.hydro}
X^{\reg{i}\reg{j}} = \frac{\omega_{k}^{\reg{j}}}{\omega_{k}^{\reg{i}}} = \frac{\sound{j}}{\sound{i}} \, .
\end{equation}
The last expression in \refb{Eq.X.hydro} is only valid in the hydrodynamic approximation, where the number of particles produced from a single step is $k$-independent. The mean number of particles in each mode $k$ is given by
\begin{equation}
\vert \Bb{1}{2} \vert^{2} ={1\over4}
\left\vert \sqrt{\sound{2}\over\sound{1}} - \sqrt{\sound{1}\over \sound{2}}  \right\vert^{2} \, .
\end{equation}

The advantage of this representation lies in the simple way that it can be extended for $m$ sudden variations in a row
\begin{equation} \label{Eq.M.1m}
\Mf{1} = \underbrace{M^{\reg{1}\reg{2}} \cdot M^{\reg{2}\reg{3}} \cdots M^{(\reg{m}-\reg{1})\reg{m}}}_{M^{\reg{1}\reg{m}}} \, \Mf{m} \, ,
\end{equation}
and each single transition matrix $M^{\reg{i}(\reg{i}+\reg{1})}$ is of the form \refb{Eq.M} evaluated at $t^{\reg{i} (\reg{i}+\reg{1})}$. The resulting matrix $M^{\reg{1}\reg{m}}$ carries the final Bogoliubov coefficients $\Ba{1}{m}$ and $\Bb{1}{m}$ for the whole chain of events. We would like to point out that the choice for the mode functions in the intermediate regimes does not have any influence on the final outcome, as can be seen in Eq.~\refb{Eq.M.1m}. However for the validity of our calculation it is necessary to choose a pair of complex conjugate mode functions.
%
\begin{figure}[htb]
 \begin{center}
 \input{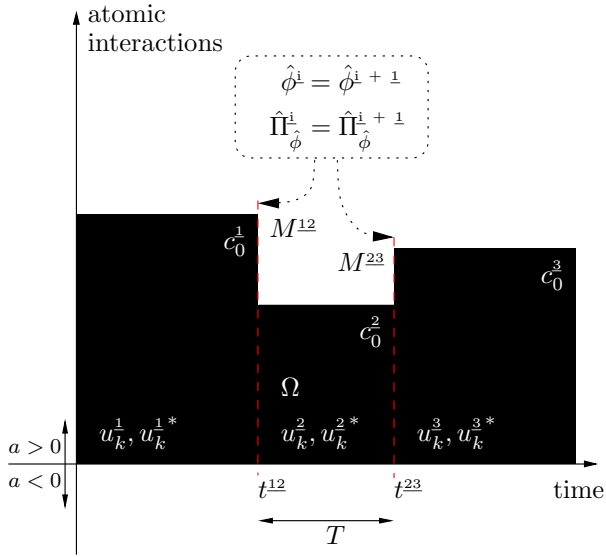}
 \caption[Figure sudden Lorentzian changes.]  {\label{Fig.Sudden.Lorentzian}
 Two sudden variations in the strength of the atomic interactions is analogous to two space-like hyper-surfaces connecting three different Minkowski spacetimes emerging form a Bose gas.}
 \end{center}
\end{figure}
%
The rest of the paper further investigates a particular scenario, that is two sudden variations in a row, see Fig.~\ref{Fig.Sudden.Lorentzian}. In this situation
\begin{eqnarray}
\Ba{1}{3} &=& \Ba{1}{2}\;\Ba{2}{3} + \Bb{1}{2}\;\Bbc{2}{3} \, , \\
\Bb{1}{3} &=& \Ba{1}{2}\;\Bb{2}{3} + \Bb{1}{2}\;\Bac{2}{3} \, .
\end{eqnarray}
To obtain the Bogoliubov coefficients for a two-step process, we
use the single-step results \refb{Eq.alpha.one.step} and
\refb{Eq.beta.one.step} to obtain:
\begin{eqnarray}
 \nonumber
&&\Ba{1}{3} = \exp\left\{i\omega_{k}^{\underline{1}}t^{\reg{1}\reg{2}}-i\omega_{k}^{\underline{3}}(t^{\reg{1}\reg{2}}+T)\right\}  \times  \\
\label{Eq.Bog.alpha.hydro}
&&  \left(\frac{ X^{\reg{1}\reg{3}}+1}{2\sqrt {X^{\reg{1}\reg{3}}}} \cos(\Omega_{k} T) - \frac {X^{\reg{1}\reg{2}}+X^{\reg{2}\reg{3}}}{i\,2\sqrt{X^{\reg{1}\reg{3}}}}
\sin ( \Omega_{k} T )\right)   \, , \;
\end{eqnarray}
and 
\begin{eqnarray}
\nonumber
&& \Bb{1}{3} = -\exp\left\{+i\omega_{k}^{\underline{1}}t_{1}+i\omega_{k}^{\underline{3}}(t_{1}+dt)\right\}   \times \\
\label{Eq.Bog.beta.hydro} 
&& \left( \frac{ X^{\reg{1}\reg{3}}-1}{2\sqrt {X^{\reg{1}\reg{3}}}} \cos(\Omega_{k} T ) +  \frac {X^{\reg{1}\reg{2}}-X^{\reg{2}\reg{3}}}{i\,2\sqrt{X^{\reg{1}\reg{3}}}} \sin ( \Omega_{k} T ) \right) \, ; \;
\end{eqnarray}
where for the finite duration intermediate region $\Reg{2}$ we
have defined the time interval
$T=t^{\reg{2}\reg{3}}-t^{\reg{1}\reg{2}}$, and the dispersion
relation $\Omega_{k}  = \omega_{k}^{\reg{2}}$. The mean number of
particles produced during this process is then given by
\begin{eqnarray}  \label{Eq.mean.number.LLL}
\vert \Bb{1}{3} \vert^{2} &=&
 \frac{(X^{\reg{1}\reg{3}}-1)^{2}}{4X^{\reg{1}\reg{3}}} \\
\nonumber
&& -\frac{1}{4}\left[X^{\reg{1}\reg{2}}- \frac{1}{X^{\reg{1}\reg{2}}}\right]\left[X^{\reg{2}\reg{3}}- \frac{1}{X^{\reg{2}\reg{3}}}\right] \sin(\Omega_{k} T)^{2}.\quad
\end{eqnarray}

It is easy to see that the mean number of particles produced in such a process oscillates between
two single step solutions of the form
\begin{equation}
N_{k} = \frac{(X-1)^{2}}{4X} \, .
\end{equation}
The upper bound replaces two up-down/ down-up steps with a single step (here given by $X=X^{\reg{1}\reg{2}}/X^{\reg{2}\reg{3}}$), and the lower bound replaces two up-down/ down-up steps with a single step  (here given by $X=X^{\reg{1}\reg{2}} X^{\reg{2}\reg{3}}$). We have illustrated the quasi-particle production in the hydrodynamic limit obtained by two-sudden steps in a row in Fig.~\ref{Fig.Hydro.LLL}.

In the following we show how to extend the calculations for finite intervals during which the microscopic atoms experience negative (attractive) interactions. For small negative values of the scattering length $a$ the condensate description continues to exist for short periods of time, hence motivates the study of Euclidean emergent geometries from a Bose gas. A sign change in the atomic interactions corresponds to a signature change in general relativity. As a first step we investigate quasi-particle production from such an event in the hydrodynamic limit.
%
%
\section{Signature change events \label{sec:Sudden.Euclidean.Variations}}
%
%
In this section we investigate the behavior of emergent spacetimes arising from a Bose gas with variations in the principal nature of the interactions; for a finite amount of time we switch to attractive atomic interactions. For short time-scales and small absolute values of the attractive $s$-wave potential, it is possible to hold on to the concept of an emergent spacetime. With the nature of the microscopic interactions switching from repulsive to attractive, the geometric hydrodynamics also changes from Lorentzian to Riemannian. This supplies us with a toy model for signature change events, which we are going to investigate next.

We first present the standard general relativity point of view, before adapting our previous calculations for the particle production due to sudden variations on finite duration Euclidean regions.
%
\subsection{Classical aspects \label{subsec:ClassicalAspects}}
Overall, we investigate manifolds that allow both Riemannian and Lorentzian regions. The latter is a generalization of Minkowski spacetimes (special relativity), while Riemannian geometries are Euclidean signature spaces including curvature (\eg, the surface of an orange). Distances in Riemannian spacetimes are positive semi-definite, while Lorentzian distances can be imaginary (time-like), zero (light-like), or positive (space-like).

In Minkowski spacetimes the signature can easily be read off as the sign of positive and negative eigenvalues of $\eta_{ab}$. From Eq.~\refb{Eq:eta.1} we get Lorentzian signature $(-,+++)$. Arbitrary curved spacetimes are locally flat, and the signature can be read off from the pattern of eigenvalues of the metric tensor $g_{ab}$ at each point on the manifold. In Lorentzian signature the ``time'' coordinate can be chosen to have a different sign from the ``spatial'' coordinates. This is in contrast to the Riemannian, or Euclidean signature $(+,+++)$, where a distinction between space and time as such does not exist \cite{Wald:1984rg}.\\

We are mainly interested in the interface between these two spacetime geometries, and we investigate the physics around a space-like hypersurface ${\tiny \sum}$ that separates the two spaces. There are two ways to be driven through a signature change, continuously or discontinuously.

We are in favour of a non-smooth signature change, to avoid degeneracies of the effective gravitational field at the surfaces of separation. For continuous signature changes the metric volume element, and hence the existence of an orthonormal frame vanish at ${\tiny \sum}$; while for discontinuous signature changes the metric volume element (hence, the orthonormal frame) is well behaved \cite{Dray:2004aa,Dray:1996dc,Ellis:1992aa}.

Thus we are allowed to transfer the connection conditions, as derived in our previous calculations for sudden changes in purely Lorentzian geometries (see Eq.~\refb{Eq.C1}, Eq.~\refb{Eq.C2}, and the discussion in the appendix), to now connect regions of Minkowskian and Euclidean spacetimes \cite{Dray:1991zz}.
\subsection{Particle production from a finite-duration Euclidian region \label{subsec:EEE}}
%
\begin{figure}[!htb]
 \begin{center}
 \input{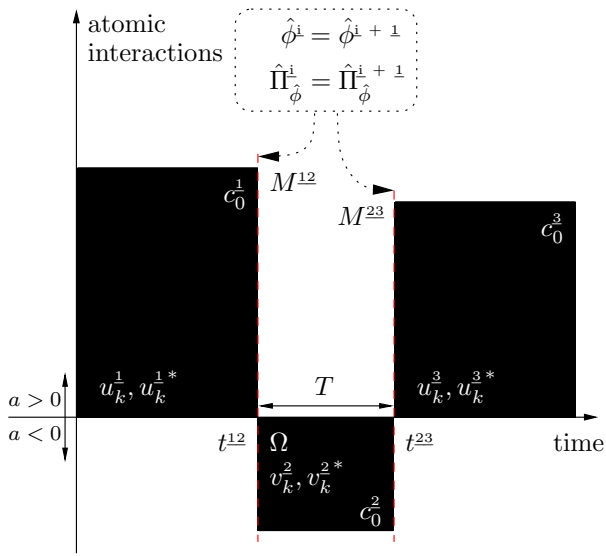}
 \caption[Figure sudden Lorentzian changes.]  {\label{Fig.Sudden.LEL}Two sudden variations in the strength of the atomic interactions are analogues to two space-like hyper-surfaces connecting three different spacetimes emerging from a Bose gas. While the initial and final geometries are Lorentzian, we allow the atomic interactions to be attractive in the intermediate region, corresponding to an Euclidean geometry.}
 \end{center}
\end{figure}
%
In the following we repeat our calculations from section \refb{sec:Sudden.Lorentzian.Variations}, but this time we choose an Euclidean geometry for region $\Reg{2}$; see figure \refb{Fig.Sudden.LEL}. As already pointed out above, the connection conditions remain the same. We further choose the mode functions in the Lorentzian regimes $\Reg{1}$, and $\Reg{3}$ to be as defined in Eq.~\refb{Eqn:Mode.Functions.Lorentzian}, where the sound speeds are well behaved; $(\sound{L})^{2}\sim U > 0$. For the intermediate Euclidean region we pick a special set of mode functions for $(\sound{E})^{2}\sim U < 0$, such that,
\begin{eqnarray}
\mfE{2}=\frac{1}{\sqrt{2 \vert \Omega_{k} \vert}} \left( \cosh(\vert \Omega_{k} \vert t) + i  \sinh(\vert \Omega_{k} \vert t) \right) \, ; \\
\mfEc{2}=\frac{1}{\sqrt{2 \vert \Omega_{k} \vert}} \left( \cosh(\vert \Omega_{k} \vert t) - i  \sinh(\vert \Omega_{k} \vert t) \right) \, .
\end{eqnarray}
The two mode functions remain a complex conjugate pair.
Here we make use of the purely imaginary dispersion relation $\Omega_{k}= i \abs{\Omega_{k}}$. In general, with the transformation matrix $S$,
\begin{equation}
S = \frac{\sqrt{i}}{2} \left(
\begin{array}{cc}
(1-i) & (1+i) \\ (1+i) & (1-i)
\end{array} \right)
\end{equation}
we can map between the two different sets of mode functions,
\begin{equation}
\MfE{2} = S \; \Mf{2} \, .
\end{equation}
Notice that $S^{-1}=S^{\dag}$ and $\det(S)=1$.

It is easy to see that $M^{\reg{1}\reg{2}} \, \Mf{2}$ transforms to $M^{\reg{1}\reg{2}} \, S \, \MfE{2}$, while $M^{\reg{2}\reg{3}} \, \Mf{3}$ transforms to $S^{-1} \, M^{\reg{2}\reg{3}} \, \Mf{3}$.
Altogether
\begin{equation}
M^{\reg{1}\reg{3}}=
M^{\reg{1}\reg{2}} S \, S^{-1} M^{\reg{2}\reg{3}}=
M^{\reg{1}\reg{2}} \, M^{\reg{2}\reg{3}}
\end{equation}
 is independent of the choice for the mode functions in the intermediate regime.

Thus we are allowed to use our previous results, replacing $\Omega{k}\rightarrow i \vert \Omega{k} \vert$, and $X^{\reg{1}\reg{2}}\rightarrow i \abs{X^{\reg{1}\reg{2}}}$ and $X^{\reg{2}\reg{3}}\rightarrow -i \abs{X^{\reg{2}\reg{3}}}$, and keeping everything else. For the mean number of particles we now obtain
\begin{eqnarray} \label{Eq.mean.number.LEL}
&&\vert \Bb{1}{3} \vert^{2} =
 \frac{(X^{\reg{1}\reg{3}}-1)^{2}}{4X^{\reg{1}\reg{3}}} \\
 \nonumber
&&\quad +\frac{1}{4}\left[\abs{X^{\reg{1}\reg{2}}}+ \frac{1}{\abs{X^{\reg{1}\reg{2}}}}\right]\left[\abs{X^{\reg{2}\reg{3}}}+ \frac{1}{\abs{X^{\reg{2}\reg{3}}}}\right] \sinh(\abs{\Omega_{k}} T)^{2}.
\end{eqnarray}
Our results are compatible with Dray \emph{et al.} \cite{Dray:2004aa, Dray:1991zz}, to the extent that the calculations and physical models overlap, but we extend the calculation for arbitrary values of the sound speeds $\sound{i}$. Furthermore we show that one can keep the calculations for purely sudden variation, as long as one picks a pair of complex conjugate mode functions in the intermediate regime. The standard general relativity calculation corresponds to $X^{\reg{1}\reg{2}}=i$, $X^{\reg{2}\reg{3}}=-i$, $X^{\reg{1}\reg{3}}=1$, so that
\begin{eqnarray} \
\vert \Bb{1}{3} \vert^{2} \to \sinh(\abs{\Omega_{k}} T)^{2}.
\end{eqnarray}
The basic reason for this tremendous simplification is that in pure general relativity (with, by definition, a single unique spacetime metric) one always has the freedom to choose coordinates such that $c=1$ in the Lorentzian region, and $c=i$ in the Euclidean region. This is a freedom we do not have in our BEC-based analogue spacetime ---
the way this shows up in our calculations is that a rapid change in the scattering length $a$ has \emph{two} effects in the condensed Bose gas: First the speed of sound is changed, modifying the ``signal cones''; and secondly the dimension-dependent conformal factor shifts by a finite amount. This second effect is absent in the traditional general relativity calculation of Dray \emph{et al.} --- for those authors it is sufficient to \emph{posit} a specific and simple change in the metric tensor $g_{ab}$ and calculate the resulting particle production. In our present situation, we first \emph{derive} a specific (dimension-independent) change in the tensor density $f^{ab}$  induced by changing the scattering length, and then \emph{derive} the corresponding (dimension-dependent) change in the metric tensor $g_{ab}$. The two situations are very closely related, but they are not quite identical.

We have plotted the mean number of particles produced in the Euclidean region in figure \ref{Fig.Hydro.LEL}. The graph shows the quasi-particle spectrum as a function of $k$. The number of particles produced depends on $\Omega_{k} T$. The longer the duration of Euclidian period, the more particles will be produced during this process.

However, there is a fundamental problem with the quasi-particle production in our effective spacetime.
Given that it is possible to connect actual condensate excitations with the calculated quasi-particle spectrum, we expect the total number of particles produced to be finite. But in both cases, for sudden variations with and without signature changes, our results imply an infinite number for the total quasi-particle production;
\begin{equation} \label{Eq.total.number.general}
N=2^{d-1}\pi \, \int_{k} dk \,k^{d-1} \; N_{k}  \, .
\end{equation}
In the next section we show how this problem resolves itself once microscopic corrections to the emergent spacetime picture are taken into account.
%
%
\section{Ultraviolet corrections \label{sec:UV-corrections}}
%
%
Up to now, we have restricted our calculations to the hydrodynamic limit, which is appropriate to describe the infrared behavior of the system. Low-energy excitations in the BEC are longitudinal phonon modes approximately propagating with the same speed; see the dispersion relation given in Eq.~\refb{Eq.disp.rel.hydro}. It is well known that so-called trans-phononic modes show a non-linear relation between excitation energy and wavelength.
In references \cite{Visser:2001ix,Weinfurtner:2006nl, Liberati:2006sj, Liberati:2006kw, Weinfurtner:2006iv} it has been pointed out that this kind of behavior might be viewed as ultraviolet corrections at the ``analogue Planck scale'', the borderline between phononic and trans-phononic modes.

Our next task is to use the eikonal approximation to include ultraviolet modes into the emergent spacetime picture.

\subsection{Rainbow geometries \label{subsec:Rainbow}}
Calculations up to this point were based on the assumption that spatial variations in the overall condensate are small. More specifically, variations in the kinetic energy of the condensate are considered to be negligible, compared to the internal potential energy of the Bosons,
\begin{equation} \label{Eq.hydrodynamic.limit}
\frac{\hbar^{2}}{2m} \, \frac{\nabla^{2} \sqrt{n_{0} + \hat n}}{\sqrt{n_{0} + \hat n}} \ll U \, .
\end{equation}
The left hand term in relation \refb{Eq.hydrodynamic.limit} is the quantum pressure term and is approximated to zero in the hydrodynamic limit.

We now keep the quantum pressure term. A straightforward computation shows that it is possible to absorb all first-order corrections into:
\begin{equation} \label{Eq.U.qp}
\mathcal{U}=U-\frac{\hbar^{2}}{4m n_{0}}\left\{
\frac{(\nabla n_{0})^{2}-(\nabla^{2}n_{0})n_{0}}{n_{0}^{2}} -\frac{\nabla n_{0}}{n_{0}^{2}}\nabla
+ \nabla^{2}
\right\} \, .
\end{equation}
That is, the effect of quantum pressure can be absorbed into an effective atom-atom interaction, see Eq.~\refb{Eq.hydro.approx}. For a uniform condensate this further simplifies to
\begin{equation}
\mathcal{U}= U-\frac{\hbar^{2}}{4m n_{0}} \nabla^{2}  \, .
\end{equation}

In order to obtain an emergent gravitational field, we apply the eikonal approximation,
\begin{equation}
\left.\mathcal{U}\right\vert_{\nabla \rightarrow -ik} \rightarrow U_{k} = U+ \frac{\hbar^{2}}{4m n_{0}} k^{2} \,.
\end{equation}
Thus also the speed of sound for ultraviolet modes has to be modified
\begin{equation} \label{Eq:sound.qp}
c_{k}^{2} = \frac{U_{k}n_{0}}{m}= c_{0}^{2}+ \epsilon_{\mathrm{qp}}^{2} k^{2} \, .
\end{equation}
Here we introduce,
\begin{equation}
\epsilon_{\mathrm{qp}} = \frac{\hbar}{2m} \, .
\end{equation}
We define the borderline between the phononic and trans-phononic modes,
\begin{equation}
\xi^{2} = \frac{1}{2} \frac{\epsilon_{\mathrm{qp}}^{2}}{c_{0}^{2}} \, ,
\end{equation}
to agree with the healing length $\xi$ of the condensate. The
healing length is the distance over which localized perturbations
in the condensate tend to smooth out \cite{Pethick:2001aa}.
Phononic excitations have wavelengths that are much larger than the healing length, $k \gg \xi$. These modes are relativistic modes in the sense of an emergent relativistic dispersion relation (see Eq.~\refb{Eq.disp.rel.hydro}). They propagate through an emergent gravitational field, given in \refb{Eq:g.1}.
Higher energy excitations around the healing scale, $k \sim \xi$, start to see deviations from the mean-field description. Such modes do not only experience collective condensate variables, they start to see the bigger picture behind the mean-field.
Consequently, they exhibit a non-relativistic dispersion relation,
\begin{equation} \label{Eq.disp.rel.qp}
\tilde{\omega}_{k} =c_{k} k=\sqrt{ c_{0}^{2} k^{2} + \epsilon_{\mathrm{qp}}^{2} k^{4}} \, .
\end{equation}
These microscopic corrections also influence the emergent spacetime, including an effective energy-dependent metric,
\begin{equation} \label{Eq:g.Rainbow}
g_{ab}^{k} = \left( \frac{c_{k}}{U_{k}/\hbar} \right)^{\frac{2}{d-1}}
\left[
\begin{array}{cccc}
-\left(c_{k}^{2}-\mathbf{v}^{2}\right) & -v_{x} & -v_{y} & -v_{z} \\
-v_{x} & 1 & 0 & 0 \\
-v_{y} & 0 & 1 & 0 \\
-v_{z} & 0 & 0 & 1
\end{array}
\right]\, ,
\end{equation}
which results in so-called rainbow geometries.

Before we continue with our program, that is to re-calculate the
mean number of particles produced including quantum pressure
effects, we would like to emphasize that all the corrections are
naturally small. For a detailed description of rainbow geometries,
and the suppression mechanism of LIV terms in Bose gases see
\cite{Liberati:2006sj,Liberati:2006kw,Weinfurtner:2006iv,Weinfurtner:2006iv,Weinfurtner:2006nl}.
%
\subsection{LIV and Particle production \label{subsec:LIV_PP}}
As mentioned above, the spacetime description only holds in the eikonal approximation. Nevertheless, a careful analysis for the particle production can be made, where the atomic interaction is a function of differential operators, see Eq.~\refb{Eq.U.qp}. For an isotropic condensate, the two different approaches will lead to the same set of equations,
\begin{equation}
\partial_{t} \left( \frac{\hbar}{U_{k}^{\reg{i}}} \; \partial_{t}  \mf{i}(t)   \right) - \frac{n_{0} \hbar}{m}\, k^{2} \; \mf{i}(t) = 0\, .
\end{equation}
The mode functions $\mfqp{i}$ are still formally represented by
Eq.~\refb{Eqn:Mode.Functions.Lorentzian}, except that we
replace $\omega^{\reg{i}}$ with $\tilde{\omega}^{\reg{i}}$, which
now includes the $\epsilon_{\mathrm{qp}}$ quantum pressure term.

In the next two sections we connect those modes over three discontinuously patched spacetime geometries, and calculate the amplification and mixing of positive and negative modes living on such a manifold.
\subsubsection{LIV and L-L-L sudden \label{subsubsec:LLL}}
We start with purely Lorentzian geometries, but allow two sudden
variations in the speed of sound. The calculations are completely
in analogy with that presented in section \refb{subsec:EEE}. Here,
the sound speeds are given by Eq.~\refb{Eq:sound.qp}, and
from the definition for the ratio $X^{\reg{i}\reg{j}}$ in the
hydrodynamic approximation, we now define:
\begin{equation} \label{Eq.X.qp}
\tilde{X}^{\reg{i}\reg{j}} = \frac{\tilde{\omega}_{k}^{\reg{j}}}{\tilde{\omega}_{k}^{\reg{i}}} = \frac{c_{k}^{\reg{j}}}{c_{k}^{\reg{i}}} \, ,
\end{equation}
as the sound-speed ratio for trans-phononic modes. Formally, the results for the Bogoliubov coefficients --- $\Ba{1}{3}$ and $\Bb{1}{3}$, given in Eqs.~\refb{Eq.Bog.alpha.hydro} and \refb{Eq.Bog.beta.hydro} --- and consequently the mean number of particles produced during this two-step process as given in Eq.~\refb{Eq.mean.number.LLL}, all remain the same. The only effort required is to replace $X^{\reg{i}\reg{j}}$ with $\tilde{X}^{\reg{i}\reg{j}}$, and $\omega_{k}^{\reg{i}}$ with $\tilde{\omega}_{k}^{\reg{i}}$.\\

We have plotted the quasi-particle spectra in figure \ref{Fig.QP.LLL}. It can be seen that the quasi-particle spectrum for the two-step process is still oscillating between the two single-step processes represented by the black and green curve. Compared to the figure on the left, where we plotted the same process in the hydrodynamic limit, see figure \ref{Fig.Hydro.LLL}, we notice that the ultraviolet particle production rapidly approaches zero.

\subsubsection{LIV and L-E-L sudden \label{subsubsec:LEL}}
Finally, we are left to analyze sudden variations from Lorentzian, to Euclidean, and back to Lorentzian spacetimes for rainbow geometries. It is interesting to notice, that in the presence of quantum pressure corrections \emph{the meaning of signature is also energy-dependent}. We distinguish between the following cases:
\begin{equation} \label{Eq:signature_qp}\mathrm{Sig.\, \Reg{2}} = \left\{
\begin{array}
{r@{\quad : \quad}l}
 \abs{k} <   {\sound{2}}/{\epsilon_{\mathrm{qp}}} & \mathrm{Euclidean} \, ; \\
 \abs{k} >  {\sound{2}}/{\epsilon_{\mathrm{qp}}} & \mathrm{Lorentzian}  \, .
\end{array} \right.
\end{equation}
We would like to emphasize that $\sound{2}$ is the sound speed in region $\Reg{2}$ for $k\rightarrow 0$.

Again, we take the previous results found in the hydrodynamic limit, see equations (\ref{Eq.Bog.alpha.hydro}, \ref{Eq.Bog.beta.hydro}, \ref{Eq.mean.number.LLL}), where we need only replace $X^{\reg{i}\reg{j}}$ with $\tilde{X}^{\reg{i}\reg{j}}$, and $\omega^{\reg{i}}$ with $\tilde{\omega}^{\reg{i}}$. \\

Altogether we expect the mean number of particles produced for $
\abs{k} > {\sound{2}}/{\epsilon_{\mathrm{qp}}} $ to be equivalent
to those from the purely Lorentzian variations, while phononic
modes, with $\abs{k} \ll {\sound{2}}/{\epsilon_{\mathrm{qp}}}$,
should experience exponential growth similar to the result from
the hydrodynamic limit. We have plotted the mean number of
particles produced for the whole event, see figure
\ref{Fig.QP.LLL}. The plots are in agreement with our predictions.

One motivation for including microscopic corrections into our hydrodynamic calculations was to solve the problem of an infinite number of particles being produced in sudden variations; see Eq.~\refb{Eq.total.number.general}. To estimate the total number of particles, we need only to focus on the region $ \abs{k} > {\sound{2}}/{\epsilon_{\mathrm{qp}}} $. It can be seen that for large $k$ the mean number of particles scales as $N_{k} \sim k^{-4}$. Therefore, for an isotropic emergent spacetime (with two or three spatial dimensions $d$), we get
\begin{equation}
N_{\leqslant k}\sim 2^{d-1}\pi \, \int dk \,k^{d-5} \sim k^{d-4} .
\end{equation}
Therefore, the total number of particles produced is finite, as expected. However, it is easy to see that the total energy emitted,
\begin{equation} \label{Eq.total.engery.general}
E=2^{d-1}\pi \, \int_{k} dk \,k^{d-1} \; N_{k} \, \omega_{k}  \sim  2^{d-1}\pi \, \int_{k} dk \, k^{d-3} \, ,
\end{equation}
is still infinite. Note, that in the ultraviolet regime one has to
use the non-linear dispersion relation $\omega_{k}\sim k^{2}$; see
Eq.~\refb{Eq.disp.rel.qp}.  We will discuss this
further in the conclusion, after we investigate a possible application of signature change events for laboratory cosmology.
%
%
\section{Particle Amplifier for Cosmological Particle Production \label{sec:amplifier}}
%
In a Euclidean-signature emergent spacetime some modes (depending on the strength of the attractive interactions) grow and decay exponentially. As pointed out in \cite{Calzetta1:2003xb} this behavior is in analogy to cosmological particle production, where super--Hubble horizon modes (\ie, modes with frequencies that are smaller than the Hubble frequency) show similar kinematics. These modes are not free to oscillate, as they get dragged along with the spacetime fabric. 
This motivated us to investigate how pre-existing condensate perturbations are influenced by the existence of a short-duration Euclidean phase. To be more specific, we would like to determine whether significant amplification of a pre-existing particle spectrum occurs from exposing it to a short finite time of attractive interactions.
This might be of interest for laboratory cosmology in Bose gases. There one of the outstanding problems involved with such experiments, as suggested in \cite{Fedichev:2006mc, Fedichev:2004on,
Barcelo:2003yk, Barcelo:2003ia, Fedichev:2004fi, Uhlmann:2005rx}, is the detectability of the quasi-particle spectrum in an emergent FRW-type universe. Due to the smallness of mode-population, common detection mechanisms would fail to measure the spectra. Recently there has been some attempt to solve this problem involving a more sophisticated detection mechanism \cite{Schutzhold:2006pv}. Another way to tackle this problem is to amplify the quasi-particle spectrum so that common detection mechanisms can be applied.
%
\begin{figure}[htb]
 \begin{center}
 \input{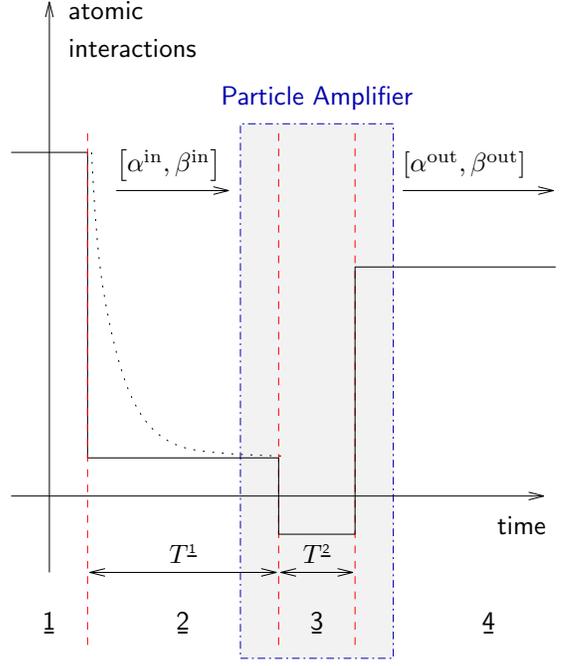}
 \caption[Figure Particle Amplifier.]  {\label{Fig.Particle.Amplifier}
The figure shows a particle production process starting from the vacuum in region $\Reg{1}$.  The bold line corresponds to particles produced in a single Lorentzian step from region $\Reg{1} \to \Reg{2}$, while the dotted line shows particles produced during region $\Reg{2}$.  Region $\Reg{3}$ depicts a subsequent amplification process of a finite time Euclidean interval.
Altogether the $\mathrm{OUT}$ signal in region $\Reg{4}$ can be viewed as an amplified $\mathrm{IN}$ signal.}
\end{center}
\end{figure}
%
\subsection{Main concept \label{Sec:Main.Concept.PA}}
In Fig.~\refb{Fig.Particle.Amplifier} we illustrate the principle of our particle amplifier idea. The particle amplifier involves a $3$-step process. Region $\Reg{2}$ will create some quasi-particle spectrum $\Bb{1}{2}=\beta^{\mathrm{IN}}$, which we wish to amplify. The Euclidean region $\Reg{3}$ is the core of our particle amplifier process with $\Bb{2}{4}=\beta^{\mathrm{PA}}$, and region $\Reg{4}$ contains the amplified quasi-particle spectrum, $\Bb{1}{4}=\beta^{\mathrm{OUT}}$. For simplicity we focus on a Lorentzian sudden step process as the source of our initial quasi-particle spectrum,
however we stress that this method is more general and applies to any quasi-particle spectrum in region $\Reg{2}$ that is of a FRW-type universe.

We would like to determine an expression for our final amplified spectrum, in terms of the input spectrum, and the spectrum we would expect from the amplifying step. We know the connection
matrix after the amplifying step can be decomposed as (see Eq.~\refb{Eq.M.1m}):
\begin{eqnarray} \nonumber
&&M^{\reg{1}\reg{4}} =M^{\reg{1}\reg{2}} \cdot  M^{\reg{2}\reg{4}} \\
&\;&=
\begin{bmatrix}\Ba{2}{4}\Ba{1}{2}+\Bbc{2}{4}\Bb{1}{2}
& \Ba{1}{2}\Bb{2}{4}+\Bb{1}{2}\Bac{2}{4}\\
\Bbc{1}{2}\Ba{2}{4}+\Bac{1}{2}\Bbc{2}{4} &
\Bbc{1}{2}\Bb{2}{4}+\Bac{1}{2}\Bac{2}{4}
\end{bmatrix} \, .
\end{eqnarray}
We now look at an expression for the final output spectrum in terms of the input spectrum and particle amplifier contribution. As we already have the Bogoliubov coefficients, this is straightforward to calculate. As $\det(M^{\reg{i}\reg{j}})=1$ it follows that:
\begin{eqnarray}
\nonumber
\vert\Bb{1}{4}\vert^{2}&=&\vert\Bb{2}{4}\vert^{2}+[2\vert\Bb{2}{4}\vert^{2}+1]\vert\Bb{1}{2}\vert^{2} \\
\label{Eq:beta.out.14.general}
&+& 2 \, \mathrm{Re}\left(\Bac{2}{4}\Bb{1}{2}\Bbc{2}{4}\Bac{1}{2}\right) \, .
\end{eqnarray}
This can be rewritten completely in terms of $\mathrm{IN}$, $\mathrm{OUT}$ and $\mathrm{PA}$ formalism,
\begin{eqnarray}
\nonumber
\vert\beta^{\mathrm{OUT}}\vert^{2}&=&\vert\beta^{\mathrm{PA}}\vert^{2}+[2\vert\beta^{\mathrm{PA}}\vert^{2}+1]\vert\beta^{\mathrm{IN}}\vert^{2} \\
\label{Eq:beta.out.IN.OUT.general}
&+& 2 \, \sin(\gamma) \; \vert\alpha^{\mathrm{PA}}\vert  \vert\beta^{\mathrm{PA}}\vert \vert\alpha^{\mathrm{IN}}\vert  \vert\beta^{\mathrm{IN}}\vert \, ,
\end{eqnarray}
where 
\begin{equation}
\sin(\gamma) = \mathrm{Re}\left(ph(\Bac{2}{4})ph(\Bb{1}{2})ph(\Bbc{2}{4})ph(\Bac{1}{2})\right) \, ,
\end{equation}
represents the relationship between the phases, $ph(...)$, of our Bogoliubov coefficients. Therefore the amplification process depends on $\gamma$, and it is necessary to find an explicit expression for it.

Although it would be possible to make further statements while keeping the analysis completely general, we now choose an explicit particle amplifier configuration and provide an example that demonstrates the parameters $\gamma$ depends on.
\subsection{Simple example \label{Sec:Simp.Ex.PA}}
For the simplest case, we look at a Euclidean amplifying step, with $X^{\reg{2}\reg{3}}=i, \, X^{\reg{3}\reg{4}}=-i,\, X^{\reg{2}\reg{4}}=1$ and write $\omega_{k}^{\underline{3}}=i|\omega_{k}^{\reg{3}}|$. This describes the scenario where the speeds of sound before and after the Euclidean step are equivalent. With these replacements the Bogoliubov coefficients for the Euclidean amplifying step can be split up as follows:
\begin{eqnarray}
\Ba{2}{4}&=&e^{i\omega_{k}^{\reg{2}}(t^{\reg{1}\reg{2}}+T^{\reg{1}})-i\omega_{k}^{\reg{4}}(t^{\reg{1}\reg{2}}+T^{\reg{1}}+T^{\reg{2}})} \,
\vert\Ba{2}{4}\vert \, , \\
\Bb{2}{4}&=&e^{+i\omega_{k}^{\reg{2}}(t^{\reg{1}\reg{2}}+T^{\reg{2}})+i\omega_{k}^{\reg{4}}(t^{\reg{1}\reg{2}}+T^{\reg{1}}+T^{\reg{2}})} \,
\vert \Bb{2}{4} \vert \, .
\end{eqnarray}
Using the notation for the Bogoliubov coefficients describing the input spectrum step to be;
\begin{eqnarray}
&\Ba{1}{2}=e^{i(\omega_{k}^{\reg{1}}-\omega_{k}^{\reg{2}})t^{\reg{1}\reg{2}}} \,
\vert\Ba{1}{2}\vert \, , \\ 
&\Bb{1}{2}=e^{i(\omega_{k}^{\reg{1}}+\omega_{k}^{\reg{2}})t^{\reg{1}\reg{2}}} \,
\vert\Bb{1}{2}\vert \, ,
\end{eqnarray} 
we then find the output spectrum after amplification becomes:
\begin{eqnarray}
\nonumber
\vert \Bb{1}{4}\vert^{2}&=&
\vert \Bb{2}{4}\vert^{2}+\left(2\vert \Bb{2}{4}\vert^{2}+1\right) \vert \Bb{1}{2}\vert \\
\label{Eq:PA.simple.example}
&&- 2 \, \sin\left(2\omega^{\reg{2}}_{k}\, T^{\reg{1}}\right)  \\  
&&  \; \times \sqrt{(\vert \Bb{1}{2}\vert^2+1)(\vert\Bb{2}{4}\vert^2+1)} \, \vert\Bb{2}{4}\vert \,  \vert \Bb{1}{2}\vert \, , \,
\nonumber
\end{eqnarray}
where we have employed the normalization; $\det(M^{\reg{i} \reg{j}})=1$.
In this particular case we found $\gamma = -2\omega^{\reg{2}}_{k}\, T^{\reg{1}}$, which therefore only depends on known parameters --- the time-intervall $T^{\reg{1}}$ and the dispersion relation $\omega^{\reg{2}}_{k}$ in region $\Reg{2}$ ---  that are tunable (for a certain range of $k$-values) throughout a BEC experiment.

Depending on $\gamma$ the number of $\mathrm{OUT}$ particles oscillates around
\begin{eqnarray}
\vert \beta^{\mathrm{OUT}}\vert^{2}&=&
\left( 1 + 2\, \vert \beta^{\mathrm{IN}}\vert^{2}  \, + \frac{ \vert \beta^{\mathrm{IN}}\vert}{ \vert \beta^{\mathrm{PA}}\vert} \right) \;  \vert \beta^{\mathrm{PA}}\vert^{2} \, ,
\label{Eq:PA.simple.example.around}
\end{eqnarray}
and thus the difference between a process with and without initial particles is given by,
\begin{eqnarray}
\vert \beta^{\mathrm{OUT}}\vert^{2}&\approx&
 2\, \vert \beta^{\mathrm{IN}}\vert^{2}   \;  \vert \beta^{\mathrm{PA}}\vert^{2} \, .
\label{Eq:PA.simple.example.around.diff}
\end{eqnarray}
In the last step we assumed that $\vert \beta^{\mathrm{IN}}\vert \ll \vert \beta^{\mathrm{PA}}\vert$, which is plausible since we expect a much larger particle production in the Euclidean region. 

Nevertheless, we have shown that for the simplest example possible the particle amplifier does indeed amplify the initial spectrum, see Eq.~\refb{Eq:PA.simple.example.around.diff}. 
However, considerably more effort is required to analyse the particle amplifier in depth.  It will be important to explore how a more realistic initial particle spectrum resulting from cosmological particle production is amplified and the efficiency of our proposed model.  The difference between the maximum possible particle production from the amplifying step ($\sin(\gamma)=-1$) and the minimum particles produced ($\sin(\gamma)=1$) might be a possible way to increase the efficiency of the particle amplifier.
In the next section we will review our results, connect with the
existing literature, and discuss problems and open questions.
%
%
\section{Conclusions \label{sec:EndHappy}}
%
%
%
\begin{figure*}[!htb]
\begin{center}
\mbox{
\subfigure[$\,$ Hydrodynamic limit; L-L-L. \label{Fig.Hydro.LLL}]{\includegraphics[width=0.45\textwidth]{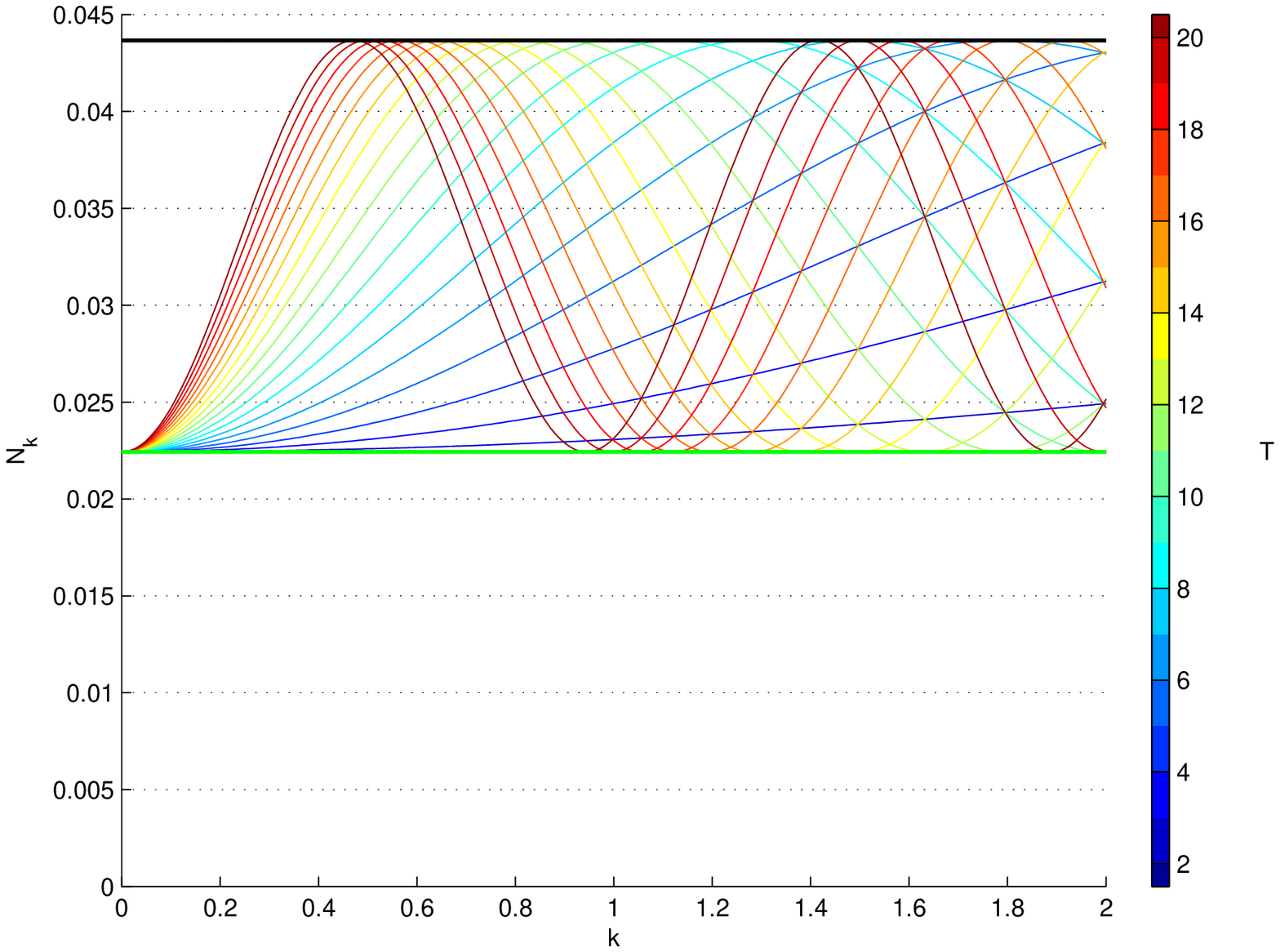}}
\hspace{0mm}
\subfigure[$\,$ Microscopic corrections; L-L-L. \label{Fig.QP.LLL}]{\includegraphics[width=0.45\textwidth]{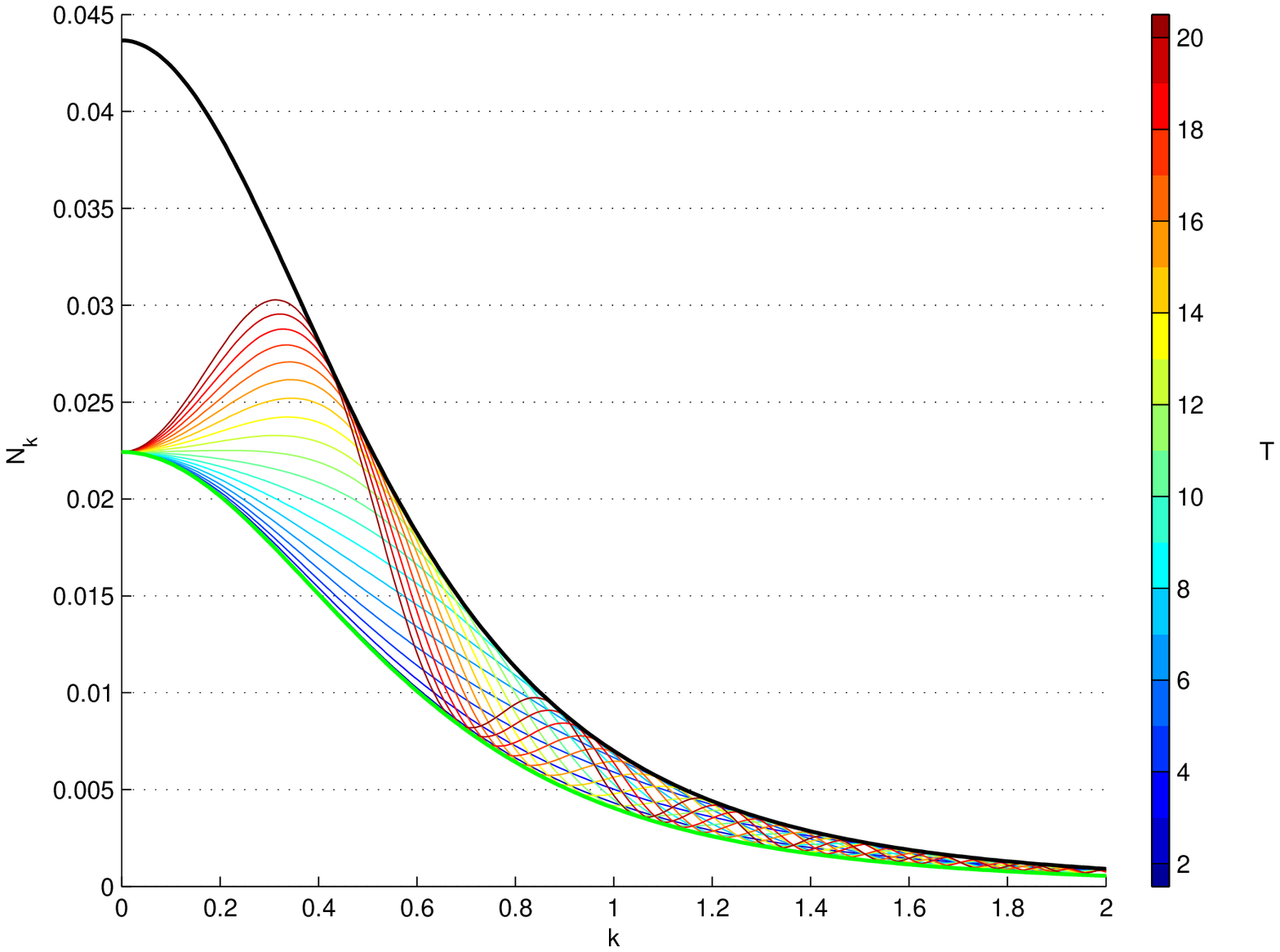}}
}
\mbox{
\subfigure[$\,$ Hydrodynamic limit; L-E-L. \label{Fig.Hydro.LEL}]{\includegraphics[width=0.45\textwidth]{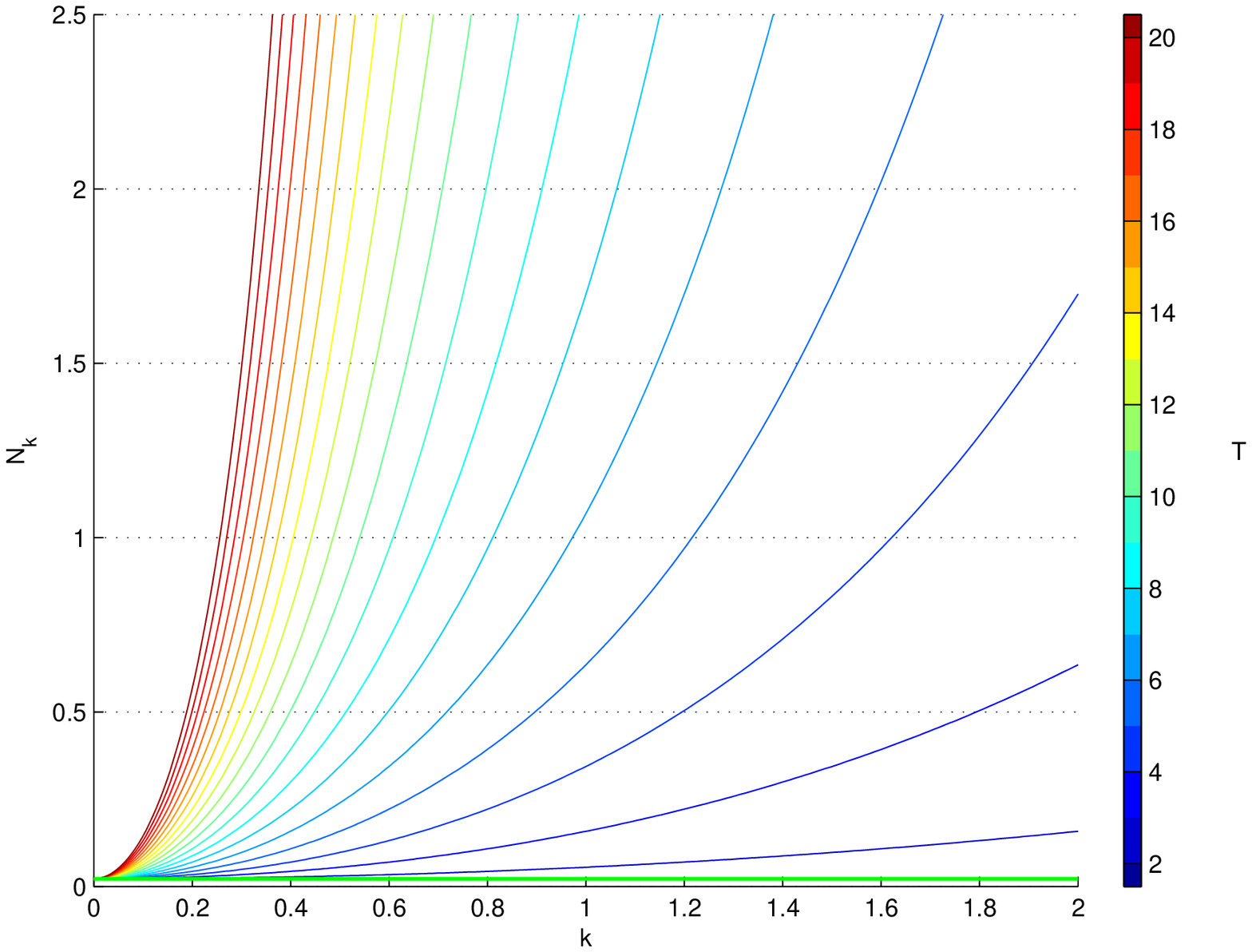}}
\hspace{0mm}
\subfigure[$\,$ Microscopic corrections; L-E-L. \label{Fig.QP.LEL}]{\includegraphics[width=0.45\textwidth]{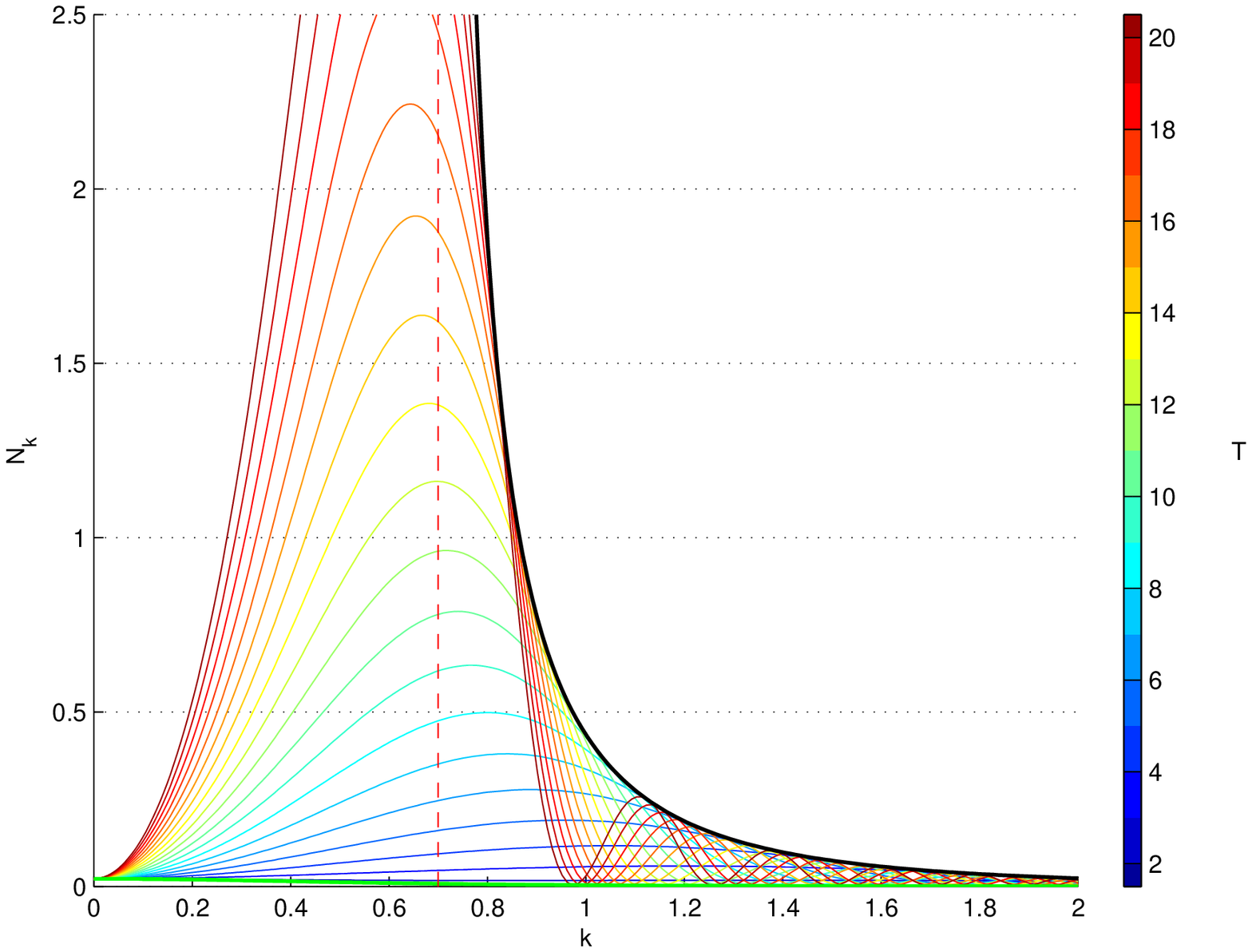}}     }
\caption{The figure displays the mean number of particles $N_{k}$ produced from two sudden variations in the strength of the atomic interactions. The different curves (color online) correspond to the time period $T$ for which the system is kept in the intermediate regime. The left column focusses on purely collective effects, while in the right column microscopic corrections are taken into account. The top two diagrams show variations between spacetimes emerging from a Bose gas with repulsive interactions ($a>0$). In the bottom row of the diagrams the intermediate regime exhibits an Euclidean geometry, where the underlying Bose gas shows attractive atomic interactions ($a<0$).}\label{fig:desitterresults}.
\end{center}
\end{figure*}
%
In the first few pages of this article we introduced the reader to
the concept of emergent spacetimes. Under certain conditions, the
microscopic theory of an ultra-cold weakly interacting Bose gas
gives way to a mean-field description, the Bose--Einstein
condensate. A new state of matter has then been formed, described
by a complex classical field. Its behavior is dominated by
collective variables, rather then individual atoms. A careful
study of the quantum excitations around the mean-field reveals the
correspondence between the condensed matter physics and quantum
field theory in curved spacetimes.  The quantum perturbations are
massless spin-zero particles, that are controlled by an emergent
geometrical/gravitational field.

The actual excitations $\delta\hat\psi$ and $\delta\hat\psi^{\dag}$ are related to perturbations in the collective Hermitian variables, the phase $\hat\phi$ and density $\hat n$ of the complex mean-field. The condensed matter description for linearized perturbations supplies commutation relations for the phase and (up to a function) its material derivatives. We were able to show that these commutators are precisely the ones for the field $\hat\phi$ and its canonical momentum $\hat\Pi_{\hat\phi}$ on the emergent curved spacetime.

The effective gravitational field defined on  the Bose gas
naturally implements a mechanism to experimentally perform
signature-change events. The signature of the emergent
gravitational field is directly correlated to the sign of the
$s$-wave scattering amplitude of the microscopic collisions. From
a gas with repulsive/ attractive interaction we expect an emergent
Lorentzian/ Riemannian geometry. The existence of Riemannian
geometries in a Bose gas is limited to short duration time
intervals and weak interactions, since back-reaction and
instability issues would otherwise destroy the condensate; see
\cite{Hu:2005wu,Calzetta1:2003xb,Calzetta:2005yk,Hu:2003}.

Inspired by recent condensed matter experiments
\cite{Donley:2001aa,Roberts:2001aa}, and theoretical work based on
them \cite{Hu:2005wu,Calzetta1:2003xb,Calzetta:2005yk,Hu:2003}, we
investigated the quasi-particle production caused by an effective
gravitational field going through a sudden but finite duration
transition to Euclidean signature. We compared the signature
changing case with sudden variation between different Lorentzian
regimes, and showed that perturbations in the Euclidean regime
experience exponential growth, while for an intermediate
Lorentzian regime the modes oscillate between single steps. These
are given by the maxima and minima configuration of the two-step
process with $T=0$. Our results are in two ways a generalization
of the existing literature, (\eg, the standard general relativity
calculations of Dray \emph{et al.} \cite{Dray:2004aa,
Dray:1991zz}):

1) For the right choice of mode functions (a complex conjugate
pair), the calculation is formally equivalent to purely Lorentzian
sudden variation between arbitrary levels. This formalism can be
applied to any $n$-step process.

2) In addition, we went beyond the hydrodynamic description, which
is only sufficient to account for the infrared behavior of the
system, and included ultraviolet modes in our analysis. This leads
to modifications in both the emergent gravitational field and the
dispersion relation. For the gravitational field we obtained (in the
eikonal limit) a momentum-dependent rainbow geometry; and a
non-relativistic dispersion relation. The modification in the
dispersion relation can be classified as being within the boost
sub-group, and is of the form as suggested by some effective field
theories \cite{Mattingly:2005aa, Liberati:2006sj, Liberati:2006kw,
Weinfurtner:2006iv, Weinfurtner:2006nl}.

In figure \refb{fig:desitterresults} we illustrated our four different results, namely sudden variations without (first row) or with (second row) signature change events, within the hydrodynamic limit (left column), and beyond it (right column). As a result, the total number of particles produced is finite. One way to explain the behavior of the trans-Planckian modes is to understand the physical significance of the healing length of the Bose--Einstein condensate. The healing scale defines a length scale over which quantum fluctuations in the condensate tend to smooth out. In some sense, it defines the smallest possible size for a condensate, $V > \xi^{d}$, see \cite{Pethick:2001aa}. Perturbations with wavelengths that are much larger than the healing length experience a nice smooth mean-field, here our emergent gravitational field. High energy modes, with wavelength at the magnitudes of the healing length start to see fingerprints of the microscopic structure. Note that, perturbations of sufficiently high energy are not driven by collective variables. This can also be seen in figure \ref{Fig.QP.LEL}: While infrared modes behave like their hydrodynamic/ relativistic neighbours to the left \ref{Fig.Hydro.LEL}, the ultraviolet modes do not notice the change in the signature of the gravitational field. They are behaving just like their upper neighbours \ref{Fig.QP.LLL} in a purely Lorentzian geometry.

Still, not all of our concerns are yet resolved. For example,
while the total number of particles produced can be made finite,
the total energy remains formally infinite even beyond the
hydrodynamic limit. This problem might be related to our choice of
ultra-high-energy description for the fundamental Bosons.
Perturbations of sufficiently high energy can excite single atoms
out of the condensate, and these are (in our description)
non-relativistic particles. Therefore, we suggest that to improve
the ultra-high-energy behaviour it might be useful to start with a
fully relativistic description for the fundamental Bosons. (For a
description of relativistic Bose--Einstein condensates see, \eg,
\cite{Bernstein:1991aa}.)

Finally, we would like to comment on the possible connection between our theoretical results and the data from the Bose-nova experiment carried out by Donley \emph{et al.} \cite{Donley:2001aa, Roberts:2001aa}. There are fundamental differences between our approach, and the Bose-nova experiment. First, we have chosen a hard-walled box as an external potential, while in the experiment a harmonic trap has been used. In \cite{Hu:2005wu,Calzetta1:2003xb,Calzetta:2005yk,Hu:2003}, Hu and Calzetta have shown that the trapping energy delays the condensate collapse for a certain amount of time. Their calculations are in good accordance with the experimental values. In our case the trapping energy is zero, and hence we do not expect such a delay.

Another fundamental difference is that we kept our gravitational
field non-degenerate (apart from the actual instant of signature
change). 
In the experiment the initial scattering length was taken
to be zero and held there for a finite time, and hence Donley
\emph{et al.} started with a partly \emph{degenerate} gravitational
field. (Partly degenerate gravitational field, because for a harmonic trap the quantum pressure term exhibits a $k$-independent contribution resulting in a position-dependent correction to the effective atomic interaction; see first term inside the curly brackets in Eq.~\refb{Eq.U.qp}.)
In the experiment they have chosen two different sequences
for the sudden variations, with two different outcomes. One, where
the scattering length has been driven from zero to attractive and
finally to large repulsive atomic interactions. The other set-up
was from zero to attractive, back to zero and then to large
repulsive atomic interactions. In both cases they detected bursts
of atoms leaving the condensate. But the bursts without the
intermediate regime of zero interactions were much stronger. These
are ``jets'' of atoms, while in the other case the bursts were
less strong. Hu and Calzetta approached this problem without
taking back-reaction effects --- of the quantum perturbations onto
the condensate --- into account. Naturally, their description
becomes less accurate with the length of the Euclidean
time-interval. We would like to emphasize the difference in the
behavior for the acoustic metric, the ``gravitational field'',
since in the first scenario it does not exhibit any finite
interval of degeneracy, while in the latter it does.

Furthermore, we would like to propose a rather different use for
our L-E-L-process, as a particle amplifier for cosmological
particle production in the laboratory. In our calculations so far
we always started from the vacuum with zero collective excitations
in the system. These calculations could easily be extended, such
as to start with a non-vacuum state, \eg, after cosmological
particle production in an emergent Friedmann--Robertson--Walker
type geometry \cite{Fedichev:2006mc, Fedichev:2004on,
Barcelo:2003yk, Barcelo:2003ia, Fedichev:2004fi, Uhlmann:2005rx}.
Recently this behavior has been studied numerically in a realistic
Bose--Einstein condensate, where the present authors used a
classical-field-method approach, see \cite{Jain:2006ki}. To obtain
an effective expanding universe, the scattering length has to
decrease as a specific function of time. After a finite expansion
time, the quasi-particle spectrum obtained --- a very small number of particles
--- somehow has to be detected. Due to its smallness this
remains a significant experimental challenge. Recently, there has
been some theoretical effort regarding this problem, see
\cite{Schutzhold:2006pv}. That author suggests a rather
complicated detection mechanism, where only perturbations of one
wavelength at a time can be detected.
In section \refb{sec:amplifier} we suggested an alternative concept, that is to \emph{amplify} the quasi-particle spectrum, \eg, in a three-step process involving a brief finite-duration Euclidean region to amplify the signal, so that common detection mechanisms might be able to detect the amplified quasi-particle spectrum.  
For a simple example, we calculated the particle spectrum after the amplification process in terms of the initial spectrum. However, for any practical application in a real BEC experiment there are considerable technical issues that need to be explored. (For example, the possibility of using a L-L-L amplifier.)

Before finishing we would like to briefly discuss the possibility
of real general relativistic signature change events. In 1992,
James Hartle and Steven Hawking proposed a signature change at
extremely early times, in the very early stages of the big bang,
when quantum gravity effects are expected to be dominant (see
reference \cite{Hartle:1983aa}). They suggested that the existence
of physical time, and hence the existence of our universe, is
associated with a signature change event from Euclidean to
Lorentzian geometry (the ``no boundary'' proposal).

In some sense it is possible to consider the reverse Bose-Nova experiment (attractive to repulsive) as the creation process for our emergent spacetime. At early times the Boson interactions are strongly attractive, such that the atoms do not show any collective/ mean-field behavior. If now the interactions experience a smooth, or non-smooth change for the atomic interactions, from attractive to repulsive, the individual atoms have to give way to collective/ mean-field variables. Therefore the existence of our emergent gravitational field might in some sense be associated with a pre-dating signature change event caused by a change in the underlying microscopic variables.
%

\begin{acknowledgments}
We wish to thank Susan Scott, Craig Savage, John Close, and Carl
Wieman for fruitful conversations and support.  We also
thank Tevian Dray for supplying a manuscript of
\cite{Dray:2004aa}.
\end{acknowledgments}
\appendix
\section{Junction conditions\label{sec:Junction_conditions}}

For the benefit of readers who may not wish to deal with
differential forms and conjugate momenta, we now present a
``low-brow'' calculation of the $\alpha$ and $\beta$ coefficients
--- for the specific physics problem we are interested in --- in
two different coordinate systems.  Both co-ordinate systems lead
to the same physics result, and both give explicit expressions for
the conserved inner product, the modes, how to normalize them, and
of course the junction conditions.

\subsection{Basics}

The master PDE that everything starts from is
\begin{equation}
\partial_t \left( - {\hbar\over U} \; \partial_t \phi \right) + {n_0\hbar\over m} \; \nabla^2 \phi = 0
\end{equation}
where $n_0$, $m$ and $\hbar$ are constant, while
\begin{equation}
U(t) = U^{\reg{1}} \;\Theta(t^{\reg{1}\reg{2}}-t) + U^{\reg{2}}
\;\Theta(t-t^{\reg{1}\reg{2}}) \, .
\end{equation}
The most basic forms of the junction conditions are then that the
field operators $ \phi$, and its conjugate momentum $(\hbar/U) \;
\partial_t  \phi$, are continuous at $t^{\reg{1}\reg{2}}$.
If these conditions are not satisfied, then there is no way that
the master PDE can be satisfied at $t^{\reg{1}\reg{2}}$ --- there
would be delta-function contributions on the LHS that would not
cancel against anything on the RHS. These two junction conditions
can be written as
\begin{equation}
[{\phi}]=0;  \qquad\quad  \left[ {\hbar\over U} \; \partial_t
\phi \right] =0.
\end{equation}
Note that we are here working in terms of physical laboratory
time, and will continue to do so until we get to the subsection
where we explain how we  could have equally well done things in
terms of ``canonical time''.
One can read the conserved inner product off by starting directly
from the PDE itself --- noting that this PDE is defined for all
time, and then considering the quantity
\begin{equation}
\label{E:inner-product}
( \phi_a, \phi_b) = \int \left( \phi_a  \; {\hbar\over U} \;
\partial_t  \phi_b -  \phi_b  \; {\hbar\over U} \; \partial_t
\phi_a \right) d^d x \, ,
\end{equation}
where $ \phi_{a}$ and $ \phi_{b}$ are solutions of the master PDE.
Then by Gauss' law (or the fundamental theorem of calculus)
\begin{eqnarray}
&&
( \phi_a, \phi_b)\vert_\mathrm{final} - ( \phi_a, \phi_b)\vert_\mathrm{initial}
\\
&&
\quad \quad = \int \partial_t
\left(  \phi_a \; {\hbar\over U} \; \partial_t  \phi_b
-  \phi_b  \; {\hbar\over U} \; \partial_t  \phi_a \right) d^{d+1} x.
\nonumber
\end{eqnarray}
Now apply Leibnitz' rule, the PDE, and  an integration by parts to
obtain
\begin{equation}
( \phi_a, \phi_b)_\mathrm{final} - ( \phi_a,
\phi_b)_\mathrm{initial} = 0.
\end{equation}
So this is the correct conserved inner product for the
Klein--Gordon-like PDE we are starting with. Note that this
conservation law, the way we have set it up holds for any and all
$\mathrm{initial}$ and $\mathrm{final}$ times, regardless of which
side of the junction they are located.

\subsection{The two non-overlapping  simple regions}
In region $\Reg{1}$ ($t<t^{\reg{1}\reg{2}}$) the master PDE
reduces to
\begin{equation}
\partial_t \left( - {\hbar\over U^{\reg{1}}} \; \partial_t  \phi \right) + {n_0\hbar\over m} \; \nabla^2  \phi = 0,
\end{equation}
and can be rearranged to
\begin{equation}
- \partial_t^2  \phi + {n_0\; U^{\reg{1}} \over m} \; \nabla^2
\phi = 0.
\end{equation}
Introducing
\begin{equation}
(c^{\reg{1}})^2 = {n_0\; U^{\reg{1}} \over m} ,
\end{equation}
this becomes
\begin{equation}
- \partial_t^2   \phi + (c^{\reg{1}})^2 \; \nabla^2  \phi = 0,
\end{equation}
with the understanding that this PDE holds only in region
$\Reg{1}$ (\ie, $t<t^{\reg{1}\reg{2}}$).
Similarly, in region $\Reg{2}$ (\ie, $t>t^{\reg{1}\reg{2}}$) the
master PDE reduces to
\begin{equation}
- \partial_t^2  \phi + (c^{\reg{2}})^2 \; \nabla^2  \phi = 0.
\end{equation}
The junction conditions are unchanged, though for convenience we
can write them in terms of $c^2= n_0 \; U/m$ as
\begin{equation}\label{juncinit}
[ \phi]=0;  \qquad\quad  \left[ {1\over c^2} \; \partial_t  \phi
\right] =0.
\end{equation}

In each individual region the field satisfies the usual flat
spacetime Klein-Gordon equation, so solutions are of the form
\begin{equation}
\phi \propto \exp( i [\omega_{k} t - \mathbf{k}  \mathbf{x}]) \, ,
\end{equation}
where $\omega_{k}$ and $\mathbf{k}$ satisfy the dispersion
relation
\begin{equation}
\omega_{k} = c \, k \, ,
\end{equation}
and we have very carefully not yet specified any normalization for
these modes. (Nor is $c$ the same in the two regions, it is either
$c^{\reg{1}}$ or $c^{\reg{2}}$ as appropriate.)

One thing we can say without further calculation is this: In view
of the junction condition $[ \phi]=0$, that is $
\phi(t^{\reg{1}\reg{2}-},\mathbf{x}) =
\phi(t^{\reg{1}\reg{2}+},\mathbf{x})$,  the spatial position
dependence of the solutions to the PDE on the two sides of the
junction must be the same --- this implies that in terms of the
plane waves above we must enforce  $\mathbf{k}$ to be the same in
regions $\Reg{1}$ and $\Reg{2}$, and therefore
\begin{eqnarray}
{\omega^{\reg{1}}_{k}\over c^{\reg{1}}} = k =
{\omega^{\reg{2}}_{k} \over c^{\reg{2}}} \, .
\end{eqnarray}

\subsection{Normalization}
Now let us compare the conserved inner product for the PDE we are
physically interested in with the inner product for the naive
Klein--Gordon equation. In particular, if $\mathrm{initial}$ is
before the transition and $\mathrm{final}$ is after the
transition, then in terms of the naive ordinary Klein--Gordon
inner product the conserved inner product for our PDE is
\begin{equation}
 ( \phi_a, \phi_b)\vert_{\mathrm{initial}} = {\hbar\over U^{\reg{1}}} \; ( \phi_a, \phi_b)_\mathrm{naive} \, ;
\end{equation}
\begin{equation}
 ( \phi_a, \phi_b)\vert_{\mathrm{final}} = {\hbar\over U^{\reg{2}}} \; ( \phi_a, \phi_b)_\mathrm{naive} \, ;
\end{equation}
where $( \phi_a, \phi_b)_\mathrm{naive} = \int \left( \phi_a  \;
\partial_t  \phi_b -  \phi_b   \; \partial_t  \phi_a \right) d^d
x$ .

This means that properly normalized modes are
\begin{equation}
\sqrt{U\over\hbar} \; {1\over\sqrt{2\omega_{k}}} \; \exp( i
[\omega_{k} t - \mathbf{k}  \mathbf{x}]),
\end{equation}
which (ignoring a trivial overall constant factor, that does not
change from region $\Reg{1}$ to region $\Reg{2}$) we might as well
write as
\begin{equation}
 {c\over\sqrt{2\omega_{k}}} \; \exp( i [\omega_{k} t - \mathbf{k}  \mathbf{x}]).
\end{equation}
In particular in regions $\Reg{1}$ and $\Reg{2}$ we want to deal
with
\begin{equation}
\mf{1}(t,\mathbf{x}) =
{c^{\reg{1}}\over\sqrt{2\omega^{\reg{1}}_{k}}} \; \exp( i
[\omega^{\reg{1}}_{k} \,t - \mathbf{k} \mathbf{x}]),
\end{equation}
and
\begin{equation}
\mf{2}(t,\mathbf{x}) =
{c^{\reg{2}}\over\sqrt{2\omega^{\reg{2}}_{k}}} \; \exp( i
[\omega^{\reg{2}}_{k}\, t - \mathbf{k}  \mathbf{x}]),
\end{equation}
respectively.

\subsection{Applying the junction conditions}
In regions $\Reg{1}$ and $\Reg{2}$ we write the solutions of the
PDE as the real parts of
\begin{eqnarray}
 \phi^{\reg{1}} &=&  A {c^{\reg{1}}\over\sqrt{2\omega^{\reg{1}}_{k}}} \; \exp( i [\omega^{\reg{1}}_{k} t - \mathbf{k}  \mathbf{x}])
\nonumber\\
&& + B  {c^{\reg{1}}_{k}\over\sqrt{2\omega^{\reg{1}}_{k}}} \;
\exp(  i [-\omega^{\reg{1}}_{k} t - \mathbf{k}  \mathbf{x}]),
\end{eqnarray}
and
\begin{eqnarray}
 \phi^{\reg{2}} &=&  C {c^{\reg{2}}\over\sqrt{2\omega^{\reg{2}}_{k}}} \; \exp( i [\omega^{\reg{2}}_{k} t - \mathbf{k}  \mathbf{x}])
\nonumber\\
&& + D  {c^{\reg{2}}\over\sqrt{2\omega^{\reg{2}}_{k}}} \; \exp(  i
[-\omega^{\reg{2}}_{k} t - \mathbf{k} \mathbf{x}]).
\end{eqnarray}
Note that the $\mathbf{x}$ dependence is the same in all four of
these terms and so quietly factors out --- this is why we asserted
that $\mathbf{k}$ had to be the same in both regions.

Applying the first junction condition $[\phi]=0$, at the
transition time $t^{\reg{1}\reg{2}}$, and using the dispersion
relation implies
\begin{eqnarray}
&&\sqrt{c^{\reg{1}}} \left[A  \; \exp( i \omega^{\reg{1}}_{k}
t^{\reg{1}\reg{2}} ) + B \; \exp( - i \omega^{\reg{1}}_{k}
t^{\reg{1}\reg{2}} ) \right]
\nonumber\\
&& = \sqrt{c^{\reg{2}}} \left[ C \; \exp( i \omega^{\reg{2}}_{k}
t^{\reg{1}\reg{2}} ) + D  \; \exp( - i \omega^{\reg{2}}_{k}
t^{\reg{1}\reg{2}} )\right]. \qquad
\end{eqnarray}

The second junction condition, $[\partial_t \phi/c^2]=0$, applied
at the transition time $t^{\reg{1}\reg{2}}$,  implies (after
factoring out all the $c$'s and $\omega$'s, and using the
dispersion relation) that
\begin{eqnarray}
&& {1\over \sqrt{c^{\reg{1}}}} \left[A  \; \exp( i
\omega^{\reg{1}}_{k} t^{\reg{1}\reg{2}} ) - B \; \exp( - i
\omega^{\reg{1}}_{k} t^{\reg{1}\reg{2}} ) \right]
\nonumber\\
&& = {1\over \sqrt{c^{\reg{2}}}}
 \left[ C \; \exp( i \omega^{\reg{2}}_{k} t^{\reg{1}\reg{2}} ) - D  \; \exp( - i \omega^{\reg{2}}_{k} t^{\reg{1}\reg{2}} )\right]. \qquad
\end{eqnarray}

These two junction conditions are enough to completely specify the
transmission matrix of $\alpha$'s and $\beta$'s. Indeed, setting
$A\to 1$, $B\to 0$, $C\to \alpha$, and $D\to \beta$ and solving
for $\alpha$ and $\beta$ we find
\begin{equation}
\alpha = {1\over2} \left( \sqrt{c^{\reg{1}}\over c^{\reg{2}}}  +
\sqrt{c^{\reg{2}}\over c^{\reg{1}}} \right) \exp( -i
[c^{\reg{2}}-c^{\reg{1}}] k t^{\reg{1}\reg{2}});
\end{equation}
\begin{equation}
\beta= {1\over2} \left( \sqrt{c^{\reg{1}}\over c^{\reg{2}}}  -
\sqrt{c^{\reg{2}}\over c^{\reg{1}}} \right) \exp( i
[c^{\reg{2}}+c^{\reg{1}}] k t^{\reg{1}\reg{2}});
\end{equation}
which is exactly the result reported in the body of the paper.

\subsection{Using ``canonical time''}

Let us now pick another time coordinate.  Let us call it
``canonical time'' and define it by
\begin{equation}
T = \int {c(t)^2\over (c_*)^2} \; dt =  \int {U(t)\over U_*} \;dt,
\end{equation}
\begin{equation}
dT= {c(t)^2\over (c_*)^2} \; dt =  {U(t)\over U_*} \;dt ,
\end{equation}
where we have introduced a convenient constant reference point
$U_*$ and used this to define a convenient constant reference
speed
\begin{equation}
(c_*)^2 = {n_0 \; U_*\over m}.
\end{equation}

The the master PDE, which was in the original laboratory time
coordinate
\begin{equation}
\partial_t \left( - {\hbar\over U} \; \partial_t \phi \right) + {n_0\hbar\over m} \; \nabla^2 \phi = 0,
\end{equation}
now becomes
\begin{equation}
-\partial_T^2 \phi+ {n_0 U_*^2\over m \; U} \; \nabla^2 \phi = 0.
\end{equation}
Recognizing that ${n_0 U_*^2/(m \; U)} = {(c_*)^4/c^2}$
and defining
\begin{equation}
c_\mathrm{eff}(T) = {(c_*)^2\over c(T)},
\end{equation}
this can be re-written simply in the form of a ``parametrically
excited oscillator''
\begin{equation}
- \partial_T^2\phi + c^2_\mathrm{eff}(T) \; \nabla^2 \phi = 0.
\end{equation}

Assuming an exponential space dependence, separation of variables
yields
\begin{equation}
\phi(t,x) = \tilde\phi(t) \; \exp(-i\mathbf{k}\mathbf{x}),
\end{equation}
so that
\begin{equation}
\partial_T^2\tilde \phi = c^2_\mathrm{eff}(T) \; k^2 \;\tilde \phi.
\end{equation}

The junction conditions, which were originally  in terms of
laboratory time (refer to equation [\ref{juncinit}])
now become, in terms of canonical time, the very simple:
\begin{equation}
[\phi]=0;  \qquad\quad  \left[ \partial_T \phi \right] =0.
\end{equation}
The conserved inner product (\ref{E:inner-product})
can now be re-written
\begin{equation}
(\phi_a,\phi_b) =  {\hbar\over U_* } \;  \int \left( \phi_a
\;\partial_T \phi_b - \phi_b  \; \partial_T \phi_a \right) d^d x,
\end{equation}
that is
\begin{equation}
(\phi_a,\phi_b) =  {\hbar\over U_* } \;
(\phi_a,\phi_b)_\mathrm{naive}.
\end{equation}
So in ``canonical time'' coordinates we have a particularly simple
PDE, elementary junction conditions, and a trivial inner product
--- the only ``tricky'' thing is that we have to use the
``effective'' sound speed $c_\mathrm{eff} = {(c_*)^2/c}$.

Now in region $\Reg{1}$ the original master PDE can be cast into
the form
\begin{equation}
- \partial_T^2\phi + {(c_*)^4\over (c^{\reg{1}})^2} \; \nabla^2
\phi = 0,
\end{equation}
while in region $\Reg{2}$
\begin{equation}
- \partial_T^2\phi + {(c_*)^4\over (c^{\reg{2}})^2} \; \nabla^2
\phi = 0.
\end{equation}

In view of the form of the conserved inner product, the normalized
modes are
\begin{equation}
{1\over \sqrt{2\omega^{\reg{1}}_{k}}} \; \exp( i
[\omega^{\reg{1}}_{k} T - \mathbf{k}  \mathbf{x}]); \qquad
\omega^{\reg{1}}_{k} = {(c_*)^2\over c^{\reg{1}}} \; k;
\end{equation}
\begin{equation}
{1\over \sqrt{2\omega^{\reg{2}}_{k}}} \; \exp( i
[\omega^{\reg{2}}_{k} T - \mathbf{k}  \mathbf{x}]); \qquad
\omega^{\reg{2}}_{k} = {(c_*)^2\over c^{\reg{2}}} \; k.
\end{equation}
Note the ``odd looking'' form of the dispersion relation --- but
this is just because $\omega^{\reg{1}}_{k}$ and
$\omega^{\reg{2}}_{k}$ are not physical frequencies --- they are
``$T$-time frequencies''.

In regions $\Reg{1}$ and $\Reg{2}$ we write the solutions of the
PDE as the real parts of
\begin{eqnarray}
\phi_{\reg{1}} &=&  A {1\over\sqrt{2\omega^{\reg{1}}_{k}}} \;
\exp( i [\omega^{\reg{1}}_{k} T - \mathbf{k}  \mathbf{x}])
\nonumber
\\
&& + B  {1\over\sqrt{2\omega^{\reg{1}}_{k}}} \; \exp(  i
[-\omega^{\reg{1}}_{k} T - \mathbf{k}  \mathbf{x}]);
\end{eqnarray}

\begin{eqnarray}
\phi_{\reg{2}} &=&  C {1\over\sqrt{2\omega^{\reg{2}}_{k}}} \;
\exp( i [\omega^{\reg{2}}_{k} T - \mathbf{k}  \mathbf{x}])
\nonumber
\\
&&
 + D  {1\over\sqrt{2\omega^{\reg{2}}_{k}}} \; \exp(  i [-\omega^{\reg{2}}_{k} T - \mathbf{k}  \mathbf{x}]).
\end{eqnarray}

The first junction condition $[\phi]=0$, after applying the
dispersion relation is
\begin{eqnarray}
&& \sqrt{c^{\reg{1}}} \left [A \; \exp( i \omega^{\reg{1}}_{k}
T^{\reg{1}\reg{2}} ) + B  \; \exp( - i \omega^{\reg{1}}_{k}
T^{\reg{1}\reg{2}} )\right]
\\
&&
 =
 \sqrt{c^{\reg{2}}} \left[ C \; \exp( i \omega^{\reg{2}}_{k} T^{\reg{1}\reg{2}} ) + D \; \exp( - i \omega^{\reg{2}}_{k} T^{\reg{1}\reg{2}} )\right].
 \nonumber
\end{eqnarray}

The second junction condition $[\partial_T \phi]=0$, after
simplifying using the dispersion relation, now leads to
\begin{eqnarray}
&&{1\over \sqrt{c^{\reg{1}}}} \left [A \; \exp( i
\omega^{\reg{1}}_{k} T^{\reg{1}\reg{2}} ) - B  \; \exp( - i
\omega^{\reg{1}}_{k} T^{\reg{1}\reg{2}} )\right]
\\
&&
 =
{1\over \sqrt{c^{\reg{2}}}} \left[ C \; \exp( i
\omega^{\reg{2}}_{k} T^{\reg{1}\reg{2}} ) - D \; \exp( -
i\omega^{\reg{2}}_{k} T^{\reg{1}\reg{2}} )\right]. \nonumber
\end{eqnarray}
Up to irrelevant phases, these are the same equations as were
derived in laboratory time.

Now setting $A\to 1$, $B\to 0$, $C\to \alpha$, and $D\to \beta$
and the solving for $\alpha$ and $\beta$ we find
\begin{equation}
\alpha = {1\over2} \left( \sqrt{c^{\reg{1}}\over c^{\reg{2}}}  +
\sqrt{c^{\reg{2}}\over c^{\reg{1}}} \right) \exp( i
[\omega^{\reg{1}}_{k}-\omega^{\reg{2}}_{k}] T^{\reg{1}\reg{2}}),
\end{equation}
\begin{equation}
\beta= {1\over2} \left( \sqrt{c^{\reg{1}}\over c^{\reg{2}}}  -
\sqrt{c^{\reg{2}}\over c^{\reg{1}}} \right) \exp( i
[\omega^{\reg{1}}_{k}+\omega^{\reg{2}}_{k}] T^{\reg{1}\reg{2}}).
\end{equation}
The phases on $\alpha$ and $\beta$ are now slightly different,
but this is not new physics --- it has to do with the phases we
picked for our ``normalized modes'' --- these phases  are now
slightly different from the laboratory time calculation. Note the
magnitudes are completely unambiguous:
\begin{equation}
|\alpha| = {1\over2} \left| \sqrt{c^{\reg{1}}\over c^{\reg{2}}}  +
\sqrt{c^{\reg{2}}\over c^{\reg{1}}} \right|;
\end{equation}
\begin{equation}
|\beta|= {1\over2} \left| \sqrt{c^{\reg{1}}\over c^{\reg{2}}}  -
\sqrt{c^{\reg{2}}\over c^{\reg{1}}} \right|.
\end{equation}
In short, this appendix has served to verify that the key
technical parts of the calculation can mathematically be carried
out in a number of different ways that ultimately lead to the same
physical result.

\end{document}